\documentclass[fleqn,usenatbib]{mnras}

\usepackage[T1]{fontenc}
\usepackage{ae,aecompl}
\usepackage{xcolor}

\usepackage{graphicx}	
\usepackage{amsmath}	
\usepackage{amssymb}	
\usepackage{booktabs}
\usepackage[nolist]{acronym}
\usepackage{array}\usepackage{makecell}
\usepackage{macros}
\usepackage{ulem}

\mathchardef\mhyphen="2D

\definecolor{acolor}{RGB}{50,205,50}
\definecolor{mcolor}{RGB}{255,10,25}
\definecolor{sarahcolor}{RGB}{167, 66, 244}

\title[3D gas abundances in FIRE MW-mass simulations]{3-D gas-phase elemental abundances across the formation histories of Milky Way-mass galaxies in the FIRE simulations: initial conditions for chemical tagging}

\author[M. A. Bellardini et al.]{
Matthew A. Bellardini,$^{1}$\thanks{E-mail: mbellardini@ucdavis.edu}
Andrew~Wetzel,$^{1}$ Sarah~R.~Loebman,$^{1,2}$\thanks{Hubble Fellow}
\newauthor Claude-Andr{\'e} Faucher-Gigu{\`e}re,$^{3}$
Xiangcheng Ma,$^{4}$
and Robert Feldmann$^{5}$
\\
$^{1}$Department of Physics \& Astronomy, University of California, Davis, One Shields Ave, Davis, CA 95616, USA\\
$^{2}$Department of Physics, University of California, Merced, 5200 Lake Road, Merced, CA 95343, USA\\
$^{3}${Department of Physics and Astronomy and CIERA, Northwestern University, 2145 Sheridan Road, Evanston, IL 60208, USA}\\
$^{4}$Department of Astronomy and Theoretical Astrophysics Center, University of California Berkeley, CA 94720, USA \\
$^{5}$Institute for Computational Science, University of Zurich, Zurich CH-8057, Switzerland\\
}

\date{Accepted XXX. Received YYY; in original form ZZZ}

\pubyear{2021}

\begin{document}
\label{firstpage}
\pagerange{\pageref{firstpage}--\pageref{lastpage}}
\maketitle

\begin{acronym}
\newacro{MW}{Milky Way}
\newacro{GMC}{giant molecular clouds}
\newacro{LG}{Local Group}
\newacro{SN}{supernova}
\newacro{MDF}{metallicity distribution function}
\newacro{ISM}{interstellar medium}
\newacro{FIRE}{Feedback In Realistic Environments}
\newacro{DM}{Dark Matter}
\newacro{MFM}{Meshless Finite Mass}
\newacro{AMR}{adaptive mesh refinement}
\end{acronym}

\begin{abstract}
We use FIRE-2 simulations to examine 3-D variations of gas-phase elemental abundances of \OH{}, \FeH{}, and \NH{} in 11 MW and M31-mass galaxies across their formation histories at $z \leq 1.5$ ($t_{\rm lookback} \leq 9.4 \Gyr$), motivated by characterizing the initial conditions of stars for chemical tagging.
Gas within $1 \kpc$ of the disk midplane is vertically homogeneous to $\lesssim 0.008 \dex$ at all $z \leq 1.5$.
We find negative radial gradients (metallicity decreases with galactocentric radius) at all times, which steepen over time from $\approx -0.01 \dpk$ at $z = 1$ ($t_{\rm lookback} = 7.8 \Gyr$) to $\approx -0.03 \dpk$ at $z = 0$, and which broadly agree with observations of the MW, M31, and nearby MW/M31-mass galaxies.
Azimuthal variations at fixed radius are typically $0.14 \dex$ at $z = 1$, reducing to $0.05 \dex$ at $z = 0$.
Thus, over time radial gradients become steeper while azimuthal variations become weaker (more homogeneous).
As a result, azimuthal variations were larger than radial variations at $z \gtrsim 0.8$ ($t_{\rm lookback} \gtrsim 6.9 \Gyr$).
Furthermore, elemental abundances are measurably homogeneous (to $\lesssim 0.05$ dex) across a radial range of $\Delta R \approx 3.5 \kpc$ at $z \gtrsim 1$ and $\Delta R \approx 1.7 \kpc$ at $z = 0$.
We also measure full distributions of elemental abundances, finding typically negatively skewed normal distributions at $z \gtrsim 1$ that evolve to typically Gaussian distributions by $z = 0$.
Our results on gas abundances inform the initial conditions for stars, including the spatial and temporal scales for applying chemical tagging to understand stellar birth in the MW.
\end{abstract}




\section{Introduction}

Many current and future observational surveys of stars across the \ac{MW} seek to unveil the \ac{MW}'s formation history in exquisite detail.
Current surveys, such as the RAdial Velocity Experiment \citep[RAVE;][]{Steinmetz06}, the Gaia-ESO survey \citep{Gilmore12}, the Large Area Multi-Object Fiber Spectroscopic Telescope \citep[LAMOST;][]{Cui12}, GALactic Archaeology with Hermes \citep[GALAH;][]{DeSilva15}, and the Apache Point Galactic Evolution Experiment \citep[APOGEE;][]{Majewski17} have measured elemental abundances of hundreds of thousands of stars. Future surveys, such as the WHT Enhanced Area Velocity Explorer \citep[WEAVE;][]{Dalton12}, the Subaru Prime Focus Spectrograph \citep[PFS;][]{Takada14}, the Sloan Digital Sky Survey V \citep[SDSS-V;][]{Kollmeier17}, the 4-metre Multi-Object Spectrograph Telescope \citep[4MOST;][]{deJong19}, and the MaunaKea Spectroscopic Explorer \citep[MSE;][]{MSE19} will increase the samples of spectroscopically observed stars into the millions. A key science driver for these surveys is `galactic archaeology': to infer the history of the MW using observations of the dynamics and elemental abundances of stars today.

Measurements of stellar dynamics can provide detailed information on the \ac{MW}'s properties and formation history, but the fundamental limitation is that stellar orbits can change over time via mergers, accretion, scattering, and other dynamical perturbations \citep[e.g.][]{Abadi03, Brook04, SB09a, Loebman11}. However, a star's atmospheric elemental abundances will not change in response to these dynamical processes, providing a key orbital-invariant `tag'.
`Chemical tagging', introduced in \citet{Freeman02}, thus provides tremendous potential to infer the formation conditions of a star of arbitrary age.

`Strong' chemical tagging represents the most fine-grained scenario, to identify stars born in the same star cluster \citep[e.g.][]{PJ20}. By contrast, `weak' chemical tagging seeks to infer the general location and time where/when a stellar population formed, for example, to associate populations of stars to certain birth regions of the galaxy \citep[e.g.][]{Wojno16, Anders17} or that accreted into the \ac{MW} from galaxy mergers \citep[e.g.][]{Ostdiek20}.

Both regimes of chemical tagging rely on sufficiently precise measurements of stellar abundances and on assumptions about the elemental homogeneity (to the measured precision) of the gas from which the stars formed.
For example, strong chemical tagging of individual star clusters relies on both the internal homogeneity of the gas cloud from which the stars formed, and on how unique the abundance patterns were in that cloud across space and time. Observational evidence of open star clusters suggests the first criterion is met \citep{DeSilva07, Ting12, Bovy16} to measurable precision.
Regarding the latter criterion, observations of the \ac{MW} and external galaxies show radial and azimuthal variations in abundances across the disk \citep[e.g.][]{SanchezMenguiano16, Molla19b, Wenger19, Kreckel20}, although more work is needed to understand these spatial variations in the context of chemical tagging.
Weak chemical tagging is subject to the same assumptions but applied to larger regions of gas across the disk (or in accreting galaxies).
For example, if all gas in the disk was measurably homogeneous in all abundances at a given time, chemical tagging would offer no spatially discriminating power.
Conversely, the limit of extreme clumpiness, in which each star cluster formed with a measurably unique abundance pattern, would in principle enabled detailed chemical tagging, but it significantly would complicate the modeling.

\textit{Thus, a key question for chemical tagging is: what are the relevant spatial scales of measurable homogeneity of stars forming at a given time, and how does this evolve across cosmic time?}. \cite{BKF10} previously explored this via a toy model, where they show all star clusters $\lesssim 10^{4} \Msun$ and a large fraction of clusters with mass below $\sim 10^{5} \Msun$ are expected to be internally homogeneous.
Progress in chemical tagging requires addressing these questions regarding stellar birth before examining the subsequent dynamical evolution of stars after they form.

Many works have examined abundance variations of stars across the \ac{MW}, generally finding a negative radial gradient in abundances for stars near the plane of the disk, which flattens or turns positive at larger heights \citep[e.g.][]{Cheng12, Boeche13, Boeche14, Anders14, Hayden14, Mikolaitis14, Anders17}. Furthermore, the \ac{MW} has an observed negative vertical gradient in stars \citep{Cheng12, Carrell12, Boeche14, Hayden14}, although this slope varies significantly between observations and depends on radius \citep{Hayden14}. Both the radial and vertical gradients vary with stellar age \citep{WL19}. Additionally, \citet{Luck06, Lemasle08, Pedicelli09} found evidence for azimuthal variations in the abundances of young stars, which may result from patchy star formation \citep{Davies09, Luck11, Genovali14}.  \citet{Nieva12} also explored the homogeneity of B-type stars in the solar neighborhood, within $500 \pc$, of the sun and found scatter on the order of $0.05 \dex$ for \OH{} which they state is comparable to gas-phase abundance scatter out to $1.5 \kpc$ of the sun.

In addition to stellar abundances, many works have characterized trends of gas-phase abundances in the \ac{MW}. Observations show that the \ac{MW} has a negative radial gradient in gas-phase abundances, with a slope that varies across the elements \citep{Arellano20} and across studies \citep[e.g.][and references therein]{Molla19a}. Furthermore, evidence persists for azimuthal variations in this radial gradient, based on HII regions \citep{Balser11, Balser15, Wenger19}.

Beyond the \ac{MW}, observations of nearby \ac{MW}-mass galaxies also show negative radial gradients in gas-phase abundances \citep[e.g.][]{Pilyugin14, SanchezMenguiano16, Belfiore17, Poetrodjojo18}. 
Furthermore, some observations show azimuthal variations \citep{Sanchez15, Vogt17, Ho17, Ho18, Kreckel19, Kreckel20, SanchezMenguiano20}, while others show no azimuthal variations within measurement uncertainty ($\lesssim 0.05 \dex$) \citep[e.g.][]{Cedres02, Zinchenko16}.

Understanding how these variations change across cosmic time is imperative for chemical-tagging models.  Currently no consensus exists, amongst observations \citep[e.g.][and references therein]{Curti20} on the redshift evolution of radial elemental abundance gradients, in part because of angular resolution limitations \citep{Yuan13}, which some works have addressed via adaptive optics and gravitational lensing \citep{Jones10, Swinbank12, Jones13}.
Furthermore, different works use different calibrators to measure abundances, which often disagree \citep{Hemler20}, further complicating our understanding of spatial variations.

\begin{figure*}
    \begin{minipage}{0.24\textwidth}
        \centering
        \includegraphics[width=.95\linewidth]{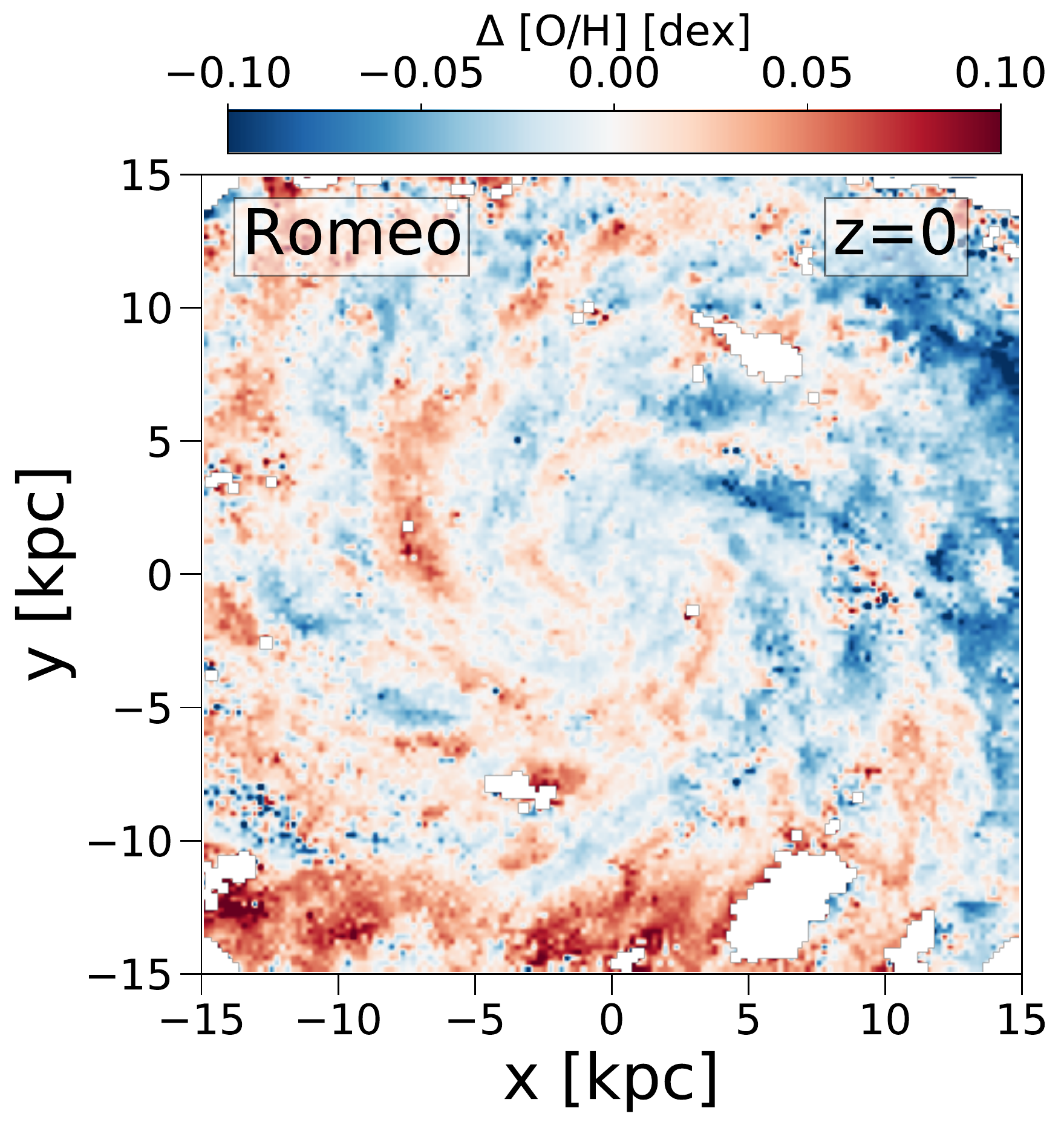}
    \end{minipage}
    \hfill
    \begin{minipage}{0.73\textwidth}
        \centering
        \includegraphics[width=.95\linewidth]{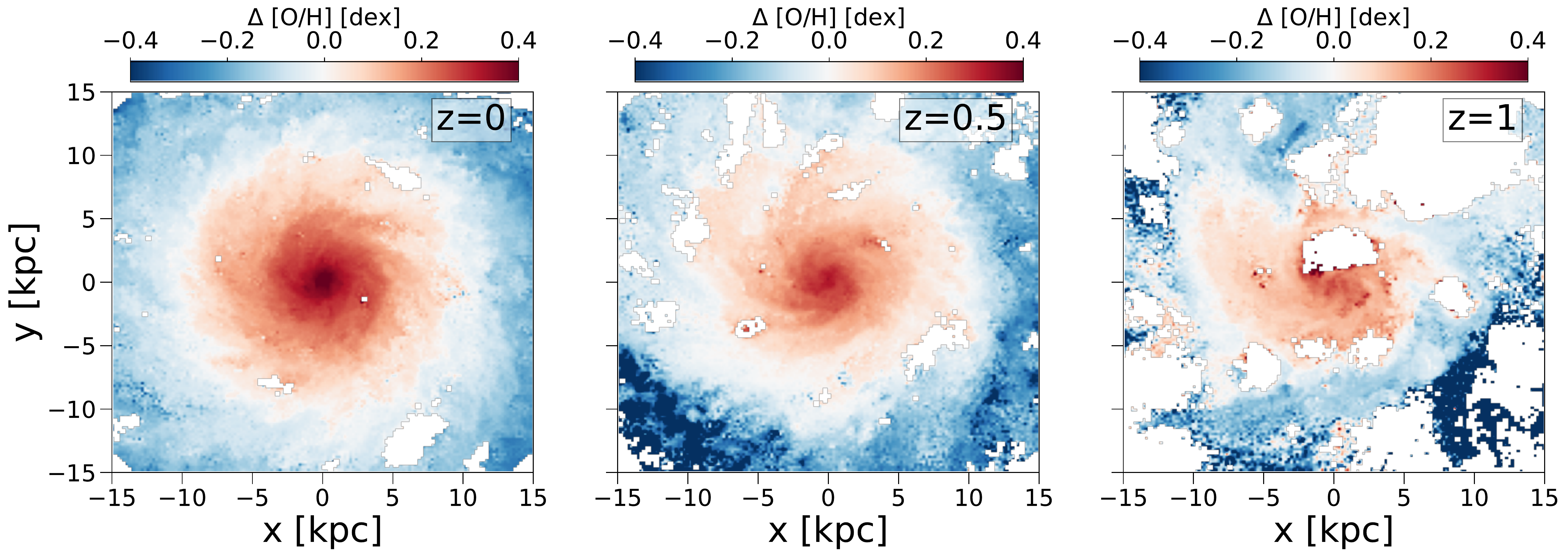}
    \end{minipage}
    \vspace{-2 mm}
    \caption{
    Face-on image of all gas within $\pm 1 \kpc$ of the galactic midplane of Romeo, one of FIRE-2 simulations that we analyze. We color-code gas by $\Delta$\OH, its deviation from the azimuthally averaged \OH. $\Delta$\FeH{} (not shown) looks nearly identical to $\Delta$\OH, to within $\lesssim 0.02 \dex$ at all times. The left panel emphasizes azimuthal variations by showing the deviation from the azimuthally averaged \OH{} \textit{at each radius}, that is, subtracting off the radial gradient, thus highlighting the enhanced enrichment correlated with spiral arms.
    The right 3 panels show the deviation from the mean \OH{} of all gas at $R \leq 15 \kpc$ at each redshift.
    White regions have highly diffuse gas in which we do not report a measured abundance.
    The radial gradient in \OH{} dominates over azimuthal variations at late times, but at early times the azimuthal variations are the most significant.
    }
\label{fig:gas_map}
\end{figure*}

Many theoretical works have used simulations to predict the spatial distribution of gas-phase abundances and their evolution.  As with observational efforts, there is no consensus for the redshift evolution of abundance gradients in theory \citep[e.g.][and references therein]{Molla19a}.
\citet{Gibson13} compared cosmological simulations with MUGS (`conservative' feedback) and MaGICC (`enhanced' feedback) run with the GASOLINE code and found that the strength of feedback in simulations is critical for the evolution of radial gradients of abundances, such that stronger feedback leads to flatter gradients at all times, while galaxies with weaker feedback have gradients that are steep at high redshift and flatten with time.
\citet{Ma17}, studying the FIRE-1 suite of cosmological simulations, found that galaxies exhibit a diverse range of radial gradients in abundances, and that these gradients can fluctuate rapidly from steep to shallow (in $\sim 100 \Myr$) at high redshift, so measurements of high-redshift gradients may not be indicative of long-term trends. They found that galaxies tend to quickly build up a negative gradient once stellar feedback is no longer sufficient to drive strong outflows of gas.
By contrast, analyzing star-forming galaxies in the TNG-50 cosmological simulation, \citet{Hemler20} found that radial gradients in galaxies are steep at high redshift and flatten with time.
Several theoretical works also have examined azimuthal variations.
\citet{Spitoni19} developed a 2-D model for abundance evolution that follows radial and azimuthal density variations in a MW-like disk and found that azimuthal residuals are strongest at early times and at large radii. Using their S2A model, \citet{Spitoni19} found azimuthal residuals in \OH{} of $\approx 0.1 \dex$ at $R = 8 \kpc$ at $t_{\rm lookback} = 11 \Gyr$ which evolve to $\approx 0.05 \dex$ at present day.
\citet{Molla19b} explored azimuthal variations in a MW-like disk for 5 models of 2-D abundance evolution and found \OH{} variations that are typically small ($0.05 - 0.1 \dex$) and dilute quickly with time.
\citet{Solar20} used young star particles as tracers of star-forming gas in the EAGLE cosmological simulation and found an average azimuthal abundance dispersion of $\approx 0.12 \dex$ at $z = 0$ in galaxies with $\Mstar = 10^9 - 10^{10.8} \Msun$.

Observations of azimuthal variations of abundances in gas in nearby \ac{MW}-mass galaxies find scatter that is comparable to observational measurement uncertainty ($\sim 0.05 \dex$) \citep{Zinchenko16, Kreckel19}. This implies that gas in galaxies is well mixed azimuthally at $z = 0$. Consequently, works modeling the abundance evolution of galaxies, which inform chemical tagging, generally assume that gas is well mixed azimuthally in the disk at all times \citep[e.g.][]{Minchev18, Molla19a, Frankel20}, such that the key spatial variation is radial. While galaxies exhibit radial gradients across a range of redshifts \citep{Queyrel12, Stott14, Wuyts16, Carton18, Patricio19, Curti20}, radial variations may not always dominate.
At early times, in particular, azimuthal variations may be more important.
Some abundance-evolution models have begun to explore both azimuthal and radial variations across time \citep[e.g.][]{Acharova13, Molla19b, Spitoni19}.
\citet{Kawata14, GKC15}, using $N$-body simulations of \ac{MW}-mass galaxies, and \citet{Baba16}, using baryonic simulations run with the SPH code ASURA-2, found that gas exhibits streaming motion along spiral arms, which could contribute to 2-D abundance variations in gas.

More detailed 2-D abundance-evolution models, which account for density variations within the disk from spiral arms and bars, result in azimuthal variations in gas-phase abundances.
\citet{Molla19b} found that arm / inter-arm abundance variations quickly dilute through interactions with spiral structure.  \citet{Spitoni19} also found that azimuthal variations dilute with time, but they found that the strength of azimuthal variations at $z = 0$ approximately agree with the observational results of \citet{Kreckel19}. The simulation analysis of \citet{GKC15, Baba16} showed that, just like stars \citep{LBK72}, gas experiences radial migration as a result of spiral structure. \citet{GKC15} found that this systematic streaming along spiral arms leads to metal-rich gas in the inner galaxy moving to larger radii and metal-poor gas in the outer galaxy moving to the inner galaxy, leading to non-homogeneous abundances at a given radius. However, they found that gas elements quickly exchange abundances after migrating, leading to small azimuthal dispersions in abundance. \citet{SanchezMenguiano20} found that azimuthal variations in abundances are stronger in galaxies with stronger bars and grand-design spirals, which supports non-axisymmetric structure driving azimuthal inhomogeneities.

In this paper we use FIRE-2 cosmological simulations of \ac{MW}/M31-mass galaxies to explore the cosmic evolution of 3-D abundance patterns of gas, as a first step towards understanding the spatial and temporal scales of applying chemical tagging in a cosmological context. Our analysis of gas represents our first step, to characterize the initial conditions for star-forming regions.
In future work, we will examine the resultant trends in stars and their dynamical evolution across time.
Here, we seek to quantify the 3-D spatial scales over which elemental abundances of gas (and thus the formation of stars) are measurably homogeneous. In Section.~\ref{sec:methods} we describe the simulations used for this analysis. In Section.~\ref{sec:results} we first explore the radial gradients and compare them against observations of the \ac{MW}, M31, and nearby \ac{MW}/M31-mass galaxies. Next we examine the cosmic evolution of radial, vertical, and azimuthal variations in gas-phase abundances, in particular, to understand which dimension dominates the spatial variations at a given time. We also examine implications of gas (in)homogeneity on current and upcoming observations of the \ac{MW}. Finally, we examine full distributions of elemental abundances.  Section.~\ref{sec:sum} we summarize the main results of the paper and provide a discussion of their implications.

\section{Methods}
\label{sec:methods}

\begin{table*}
\caption{
Properties of the stellar and gas disks of our simulated \ac{MW}/M31-mass galaxies $z = 0$. The first column lists the name of the galaxy: `m12' indicates isolated galaxies from the \textit{Latte} suite, while the other galaxies are \ac{LG} analogues from the ELVIS on FIRE suite. $R_{25}$ and $R_{50}$ is the radius where the cumulative mass of the disk is $25$\% and $50$\%, respectively, within a height $\pm 3 \kpc$ out of the midplane, relative to the total stellar/gas mass of the disk within $20$ kpc. We fit $R_{90}$ and $Z_{90}$ as the radius and height where the cumulative mass of the stellar/gas disk are $90$\% of the total mass of stars/gas within a sphere of $20 \kpc$. $M_{90}$ is the total stellar/gas mass contained within both $R_{90}$ and $Z_{90}$. The gas fraction, $f^{\rm gas}_{90}$, is the ratio of gas mass to total baryonic mass within $R^{\rm star}_{90}$ and $Z^{\rm star}_{90}$.
}
\begin{tabular}{@{}l|ccccc|ccccc|c@{}}
\toprule
\thead{simulation} & 
\thead{M$^{\rm star}_{90}$\\ $[10^{10}$ M$_{\odot}]$} & 
\thead{R$^{\rm star}_{25}$\\ $[\kpc]$} & 
\thead{R$^{\rm star}_{50}$\\ $[\kpc]$} & 
\thead{R$^{\rm star}_{90}$\\ $[\kpc]$} & 
\thead{Z$^{\rm star}_{90}$\\ $[\kpc]$} &
\thead{M$^{\rm gas}_{90}$\\ $[10^{10}$ M$_{\odot}]$} & 
\thead{R$^{\rm gas}_{25}$\\ $[\kpc]$} & 
\thead{R$^{\rm gas}_{50}$\\ $[\kpc]$} & 
\thead{R$^{\rm gas}_{90}$\\ $[\kpc]$} & 
\thead{Z$^{\rm gas}_{90}$\\ $[\kpc]$} &
\thead{f$^{\rm gas}_{90}$\\ ($< R^{\rm star}_{90}$, $< Z^{\rm star}_{90})$}\\ \midrule
m12m & 10.0 & 1.9 & 4.3 & 11.6 & 2.3 & 2.1 & 6.6 & 10.3 & 15.0 & 1.2 & 0.13 \\
Romulus & 8.0 & 1.2 & 3.2 & 12.9 & 2.4 & 2.7 & 9.0 & 13.1 & 18.3 & 2.3 & 0.16 \\
m12b & 7.3 & 1.0 & 2.2 & 9.0 & 1.8 & 1.7 & 6.4 & 9.6 & 15.0 & 1.5 & 0.11 \\
m12f & 6.9 & 1.2 & 2.9 & 11.8 & 2.1 & 2.3 & 8.7 & 12.6 & 17.8 & 2.4 & 0.14 \\
Thelma & 6.3 & 6.3 & 3.4 & 11.2 & 3.2 & 2.6 & 7.5 & 12.1 & 17.6 & 3.1 & 0.16 \\
Romeo & 5.9 & 1.6 & 3.6 & 12.4 & 1.9 & 1.8 & 8.0 & 12.2 & 18.1 & 1.5 & 0.16 \\
m12i & 5.3 & 1.1 & 2.6 & 9.8 & 2.3 & 1.7 & 7.1 & 10.2 & 16.7 & 1.7 &  0.15 \\
m12c & 5.1 & 1.3 & 2.9 & 9.1 & 2.0 & 1.5 & 5.4 & 8.4 & 14.6 & 2.4 & 0.15 \\
Remus & 4.0 & 1.2 & 2.9 & 11.0 & 2.2 & 1.5 &7.4 & 11.8 & 18.0 & 1.8 & 0.22 \\
Juliet & 3.3 & 0.8 & 1.8 & 8.1 & 2.2 & 1.5 & 6.9 & 11.3 & 18.6 & 3.1 & 0.14 \\
Louise & 2.3 & 1.2 & 2.8 & 11.2 & 2.2 & 1.4 & 8.0 & 12.6 & 18.5 & 2.0 & 0.34 \\
\bottomrule
\end{tabular}
\label{table:scale_lengths}
\end{table*}

\subsection{FIRE-2 Simulations}

We use a suite of \ac{MW}/M31-mass cosmological zoom-in simulations from the \ac{FIRE} project\footnote{FIRE project web site: \href{http://fire.northwestern.edu}{http://fire.northwestern.edu}} \citep{Hopkins18}. We ran these simulations using the \ac{FIRE}-2 numerical implementations of fluid dynamics, star formation, and stellar feedback. These simulations use the Lagrangian \ac{MFM} hydrodynamics method in \textsc{Gizmo} \citep{Hopkins15}. The \ac{FIRE}-2 model incorporates physically motivated metallicity-dependent radiative heating and cooling processes for gas such as free-free, photoionization and recombination, Compton, photo-electric and dust collisional, cosmic ray, molecular, metal-line, and fine structure processes, accounting for $11$ elements (H, He, C, N, O, Ne, Mg, Si, S, Ca, Fe) across a temperature range of $10 - 10^{10} \K$. The simulations also include a spatially uniform, redshift-dependent UV background from \cite{Faucher09}.
In calculating metallicities throughout this paper, we scale elemental abundances to the solar values in \citet{Asplund09}.

Star particles form out of gas that is self-gravitating, Jeans-unstable, cold ($T < 10^{4} \K$), dense ($n > 1000 \cm^{-3}$), and molecular \citep[following][]{KG11}. Each star particle inherits the mass and elemental abundances of its progenitor gas and represents a single stellar population, assuming a \citet{Kroupa01} stellar initial mass function. \ac{FIRE}-2 evolves star particles along standard stellar population models from e.g. STARBURST99 v7.0 \citep{Leitherer99}, including time-resolved stellar feedback from core-collapse and Ia supernovae, continuous mass loss, radiation pressure, photoionization, and photo-electric heating.
\ac{FIRE}-2 uses rates of core-collapse and Ia supernovae from STARBURST99 \citep{Leitherer99} and \citet{Mannucci06}, respectively. The nucleosynthetic yields follow \citet{Nomoto06} for core-collapse and \citet{Iwamoto99} for Ia supernovae. Stellar wind yields, sourced primarily from O, B, and AGB stars, are from the combination of models from \citet{vandenHoek97, Marigo01, Izzard04}, synthesized in \citet{Wiersma09}.

Critical for this work, these \ac{FIRE}-2 simulations also explicitly model the sub-grid diffusion/mixing of elemental abundances in gas that occurs via unresolved turbulent eddies \citep{Su17, Escala18, Hopkins18}. In effect, this smooths abundance variations between gas elements, assuming that sub-grid mixing is dominated by the largest unresolved eddies. \citet{Escala18} showed that incorporating this sub-grid model is a necessity to match observed distributions of stellar metallicities.
We explore the robustness of our results to variations in the strength of the mixing/diffusion coefficient in Appendix~\ref{sec:diffusion_coefficient_test}.

All simulations assume flat $\Lambda$CDM cosmologies with parameters broadly consistent with the \citet{Planck18}: $h = 0.68 - 0.71$, $\Omega_{\Lambda} = 0.69 - 0.734$, $\Omega_{\rm m} = 0.266 - 0.31$, $\Omega_{\rm b} = 0.0455 - 0.048$, $\sigma_{8} = 0.801 - 0.82$ and $n_{s} = 0.961 - 0.97$.  For each simulation we generated cosmological zoom-in initial conditions embedded within cosmological boxes of length $70.4 - 172 \Mpc$ at $z \approx 99$ using the code MUSIC \citep{Hahn11}. We saved $600$ snapshots from $z = 99$ to $0$, with typical time spacing $\lesssim 25 \Myr$.

We examine $11$ \ac{MW}/M31-mass galaxies from 2 suites of simulations. We select only galaxies with a stellar mass within a factor of $\approx 2$ of the \ac{MW}, $\approx 5 \times 10^{10} \Msun$ \citep{BG16}. 5 of our galaxies are from the \textit{Latte} suite of isolated individual \ac{MW}/M31-mass halos, introduced in \cite{Wetzel16}. (We exclude m12w, because it has an unusually compact gas disk at $z = 0$, with $R^{\rm gas}_{90} = 7.4 \kpc$). \textit{Latte} galaxies have halo masses $M_{\rm 200m} = 1 - 2 \times 10^{12} \Msun$, for which $M_{\rm 200m}$ refers to the total mass within the radius containing $200$ times the mean matter density of the Universe. These simulations have \ac{DM} particle masses of $3.5 \times 10^{4}\Msun$ and initial baryonic particle masses of $7070 \Msun$ (though because of stellar mass loss, star particles at $z = 0$ typically have masses of $\sim 5000 \Msun$). Star and \ac{DM} particles have fixed gravitational force softening lengths of $4$ and $40 \pc$ (Plummer equivalent), comoving at $z > 9$ and physical thereafter. Gas elements have adaptive gas smoothing and gravitational force softening lengths that reach a minimum of $1 \pc$.
We also include 6 galaxies from the `ELVIS on FIRE' suite \citep{Garrison-Kimmel14, Garrison-Kimmel19}. These simulate \ac{LG}-like \ac{MW}$+$M31 pairs. ELVIS hosts have halo masses $M_{\rm 200m} = 1 - 3 \times 10^{12} \Msun$, with $\approx 2 \times$ better mass resolution than the \textit{Latte} suite.

In general, we find few systematic differences in any of our results for the isolated galaxies versus the \ac{LG}-like pairs, the only notable difference being the relative strength of azimuthal scatter to radial gradient strength at large radius and high $z$, so we combine these suites in all of our results.

\section{Results}
\label{sec:results}

Fig.~\ref{fig:gas_map} shows face-on images of the gas disk of one of our simulations, Romeo, at several redshifts.
We color-code gas by its variation in \OH, to visualize key trends that we explore in this work. We do not show results for \FeH{}, because they are qualitatively consistent with \OH.
The left panel shows the deviation of the local \OH{} from the mean value \textit{at each radius} for radial bins of width $200 \pc$ at $z = 0$, that is, we subtract off the overall radial gradient.
This highlights the variations along the azimuthal direction at each radius, showing enhancement in \OH{} along spiral structure (Orr et al. in prep. will present a detailed analysis of metallicity enhancements along spiral arms).

The right $3$ panels show the difference between the local \OH{} and the mean \OH{} across all gas at $R \leq 15 \kpc$ and within $\pm 1 \kpc$ of the galactic midplane at each redshift.
This highlights the importance of both radial and azimuthal abundance variations in gas. At late times, the gas disk shows a clear negative radial gradient that is much stronger than the azimuthal variations. However, at earlier times, the gas disk is azimuthally more asymmetric, including cavities from local star-forming and feedback regions.
A radial gradient is less pronounced. As we will show, at $z \gtrsim 0.8$ ($t_{\rm lookback} \gtrsim 6.9 \Gyr$) the azimuthal variations in abundance at a given radius are typically larger than the radial change across the disk.

Because the absolute normalization of any elemental abundance in our simulations is uncertain, given uncertainties in underlying nucleosynthetic rate and yield models, throughout this paper we focus on \textit{relative} abundance variations, including spatial variations, evolution, and the shapes of abundance distributions rather than absolute normalizations of abundances.  

As Fig.~\ref{fig:redshift0_radial_summary} shows, gas \OH{} in our simulations is super-solar at all radii (out to $15 \kpc$).  Over comparable radial ranges,  our hosts have an average gas-phase \OH{} $\approx 0.56 \dex$ larger than what is observed in the MW \citep{Fernandez17, Wenger19, Arellano20} and M31 \citep{Zurita12, Sanders12}.  Most likely the primary cause of this high normalization is our modeling of core-collapse supernovae in FIRE-2. We assume all core-collapse supernovae to have identical (IMF-averaged) yields, but in reality, different mass progenitors produce different yields, as \citet{Muley2020} explored in the context of FIRE-2.

Furthermore, in FIRE-2 we assume nucleosynthetic yields for core-collapse supernovae from \citet{Nomoto06}, but more recent compilations \citep{Nomoto13, Pignatari16, Sukhbold16, Limongi18}   suggest $\sim 2 \times$ lower O yields (IMF-averaged), as we will show in detail with our next-generation FIRE model (Hopkins in prep.).
In that work, we also will show that the overall stellar-wind mass-loss rates are likely $\sim 2 \times$ smaller than in FIRE-2, which further contributes to lower \OH{}.
Finally, as we will explore in Gandhi et al. (in prep.), our assumed supernovae Ia rates in FIRE-2 \citep{Mannucci06} are $\sim 2 \times$ lower than more recent Ia rate constraints \citep[e.g.][]{Maoz17}.
Thus, these updates will account for up to $\sim 0.3 - 0.4$ dex lower normalization in our predicted \OH{}, and up to $0.4 - 0.5$ dex lower normalization in our \OFe{}.

\subsection{Radial profiles at z = 0}

\begin{figure}
	\includegraphics[width = 0.97 \columnwidth]{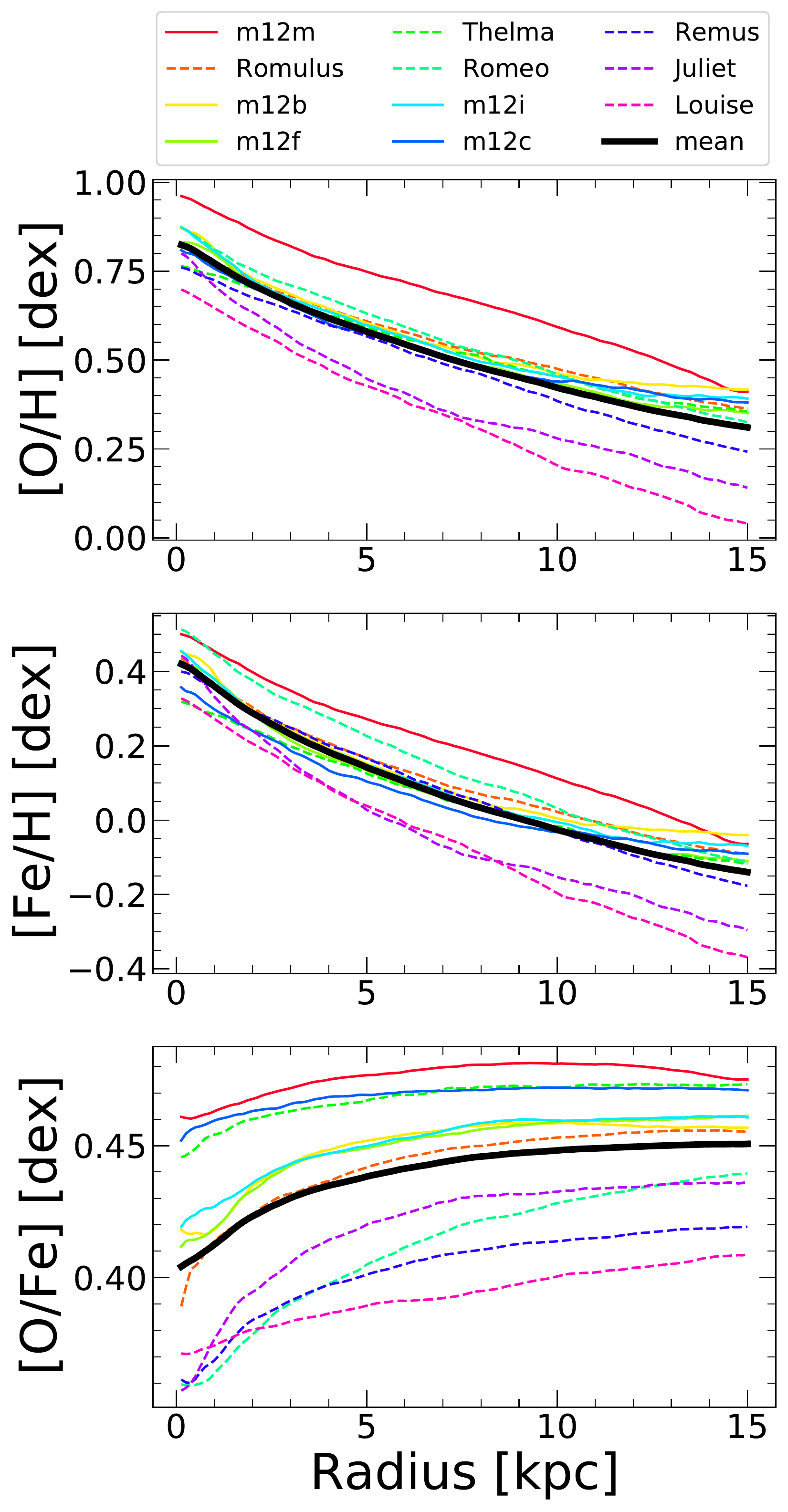}
	\vspace{-3 mm}
    \caption{
    Radial profiles of gas-phase elemental abundances within the disks of our $11$ galaxies at $z = 0$, listed by decreasing stellar mass. Each simulation includes all gas, averaging across $3$ snapshots ($\approx 50 \Myr$) within a disk height of $\pm 1 \kpc$. The black line shows the mean across all hosts. The normalization of the radial abundances scales with the stellar mass of the galaxy (see Table:~\ref{table:scale_lengths}). For \OH{} and \FeH{}, the profiles exhibit negative radial gradients, with a mean change across $0 - 15 \kpc$ of $\sim 0.51 \dex$ for \OH{} and $\sim 0.56 \dex$ for \FeH. The \OFe{} profile is approximately flat at radii $\gtrsim 4 \kpc$, indicating enrichment dominated by core-collapse supernovae, but it exhibits a small positive gradient in the inner galaxies, from the increasing importance of Fe from Ia supernovae.
    }
    \label{fig:redshift0_radial_summary}
\end{figure}

\begin{figure*}
	\includegraphics[width = \linewidth]{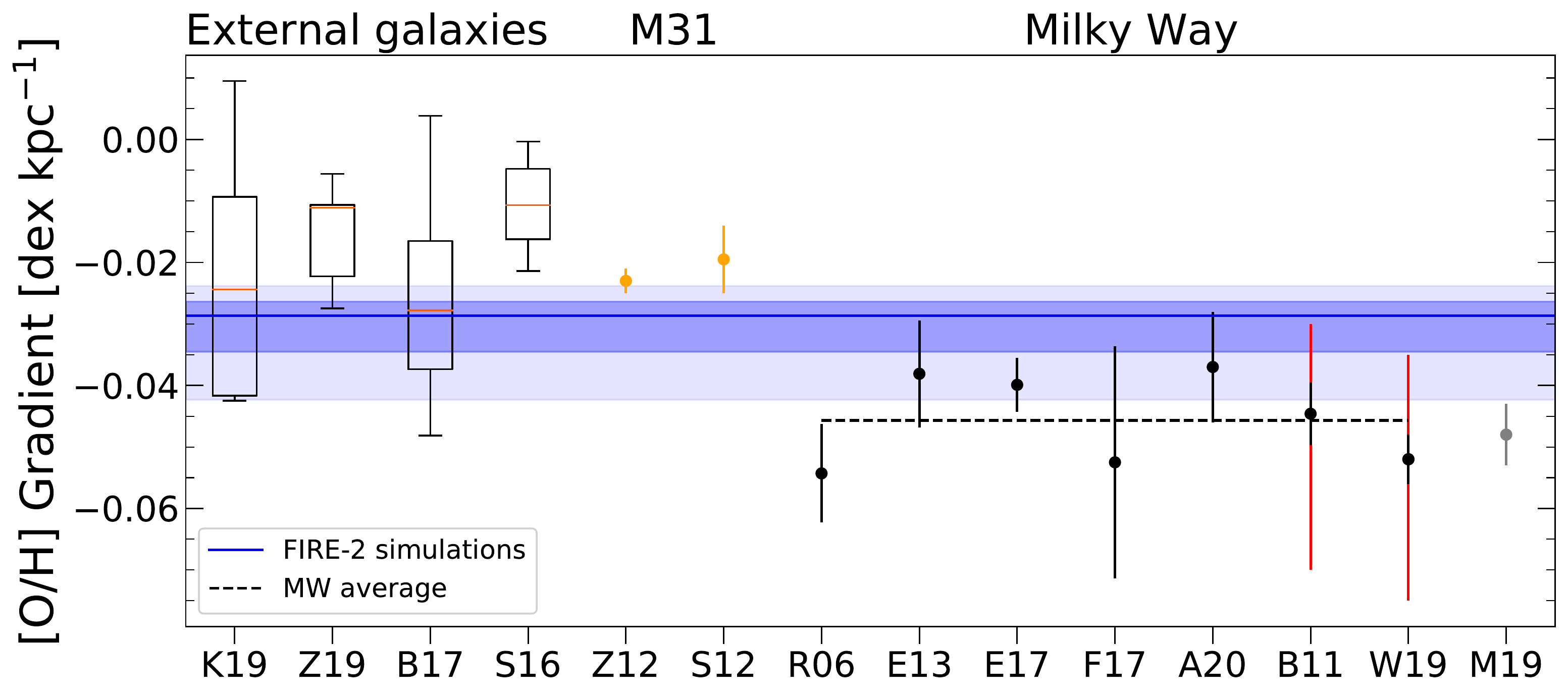}
	\vspace{-5 mm}
    \caption{
    Radial gradients in gas-phase \OH{} across our $11$ galaxies and observed in the \ac{MW}, M31, and in nearby \ac{MW}-mass galaxies. The blue horizontal line shows the median across our $11$ galaxies, with the dark shaded region showing the $68$th percentile and the light shaded region the full distribution. We also show observations of radial gradients in external galaxies, from \citet[][K19]{Kreckel19}, \citet[][Z19]{Zinchenko19}, \citet[][B17]{Belfiore17}, and \citet[][S16]{SanchezMenguiano16}, via box-and-whisker, where the box displays the $68$th percentile, the whiskers display the full distribution, and the orange horizontal line is the median. Orange circles show observed abundance gradients for M31 derived from HII regions by \citet[][Z12]{Zurita12} and \citet[][S12]{Sanders12}. Black circles show observed abundance gradients for the \ac{MW} derived from HII regions from \citet[][R06]{Rudolph06}, \citet[][B11]{Balser11}, \citet[][E13]{Esteban13}, \citet[][E17]{Esteban17}, \citet[][F17]{Fernandez17}, \citet[][W19]{Wenger19}, and \citet[][A20]{Arellano20}. \citet[][M19]{Molla19a}, shown in grey, is the gradient derived from a compilation of the data from R06, B11, E13, E17, and F17. We show uncertainties for all points. For W19 and B11 the red shows the variation in gradient observed by looking along different azimuths.  The dashed line shows the best-fit \ac{MW} gradient ($-0.046 \dpk$), based on the gradients the observations presented here (excluding M19). The median \OH{} gradient across our galaxies is $-0.028 \dpk$ and the standard deviation is $0.005 \dpk$, in agreement with K19 and B17 to within $1 \sigma$, and with Z19 to within $2 \sigma$, but not in agreement with S16. Our simulations also agree with observations of M31 and some of the \ac{MW} observations.
    }
    \label{fig:OonH_z0_sim_v_obs_grad}
\end{figure*}

First we examine the radial profiles of \OH, \FeH, and \OFe{} in gas for all $11$ galaxies at $z = 0$. We time-average each galaxy's profile across $\sim 50 \Myr$ by stacking $3$ snapshots to reduce short-time fluctuations.
We present all results in physical radii; in Appendix~\ref{sec:scale_lengths}, we examine these trends scaling to various galactic scale radii, finding that the host-to-host scatter in our suite is minimized when examining gradients in physical units.
These profiles contain all gas within a vertical height $Z \pm 1 \kpc$ from the disk and we use radial bins of width $0.25 \kpc$. We calculate the mass-weighted mean of the gas-phase abundance in each bin. We show profiles out to $R = 15\kpc$; our gas disks generally extend beyond this radius, but because our primary motivation is chemical tagging, we examine only regions with significant star formation (see Table~\ref{table:scale_lengths}).

Fig.~\ref{fig:redshift0_radial_summary} (top 2 panels) shows that \OH{} and \FeH{} decrease monotonically with radius. The mean gradient is $\approx -0.03 \dpk$ for both \OH{} and \FeH{}, and the mean change in abundance from $0 - 15 \kpc$ is $\approx 0.51 \dex$ for \OH{} and $\approx 0.56 \dex$ for \FeH. These negative gradients in gas reflect the decreasing ratio of stars (the sources of enrichment) to gas towards the outer disk, and show that these gas disks are not radially well mixed at $z = 0$.

Across our 11 galaxies, the host-to-host standard deviation is $\approx 0.09 \dex$ for \OH{} and $\approx 0.07 \dex$ for \FeH{}.
The legend of Fig.~\ref{fig:redshift0_radial_summary} lists the host galaxies in decreasing order of stellar mass, highlighting that the abundance at a given radius correlates strongly with the galaxy's mass.  Table.~\ref{table:scale_lengths} shows that stellar mass drops by a factor of $\sim 4$ from m12m to Louise; given the slope of the gas-phase mass-metallicity relation from \citet{Ma16}, $\approx 0.4 \dex$, the scatter in \OH{} normalization for our mass range should be $\approx 0.24 \dex$, almost exactly that in Fig.~\ref{fig:redshift0_radial_summary}.
In other words, the scatter across our suite primarily reflects the mass-metallicity relation \citep[see][]{Lequeux79, Tremonti04, Mannucci10, Andrews13}.

In Fig.~\ref{fig:redshift0_radial_summary}, dashed lines show the \ac{LG}-like hosts, and while they show typically lower abundance at a fixed radius than the isolated hosts, this is because they have somewhat lower stellar mass on average.
We find no systematic differences between LG-like and isolated hosts beyond this, despite the fact that the LG-like hosts form their in-situ stars systematically earlier than the isolated hosts \citep{Santistevan20}.
Thus, this difference in formation history does not imprint itself on gas-phase abundances at $z \leq 1.5$.
As a result, we will combine these samples in all subsequent analyses.

Fig.~\ref{fig:redshift0_radial_summary} (bottom panel) shows profiles for \OFe{}, which are nearly flat at all radii.
The mean change in \OFe{} from $0 - 15 \kpc$ is $\approx -0.046 \dex$.
\OFe{} shows the strongest (positive) gradient in the inner $\approx 4 \kpc$, highlighting the increasing importance of enrichment from (more delayed) Ia supernovae towards the galactic center, which underwent the longest period of enrichment. However, the outer disk, beyond $\approx 4 \kpc$, reflects relatively similar enrichment from core-collapse and Ia supernovae at each radius.
We find a host-to-host standard deviation of $\approx 0.027 \dex$ for \OFe.  We measure this for \MgFe{} (not shown here) as another tracer of core-collapse vs Ia supernovae enrichment.  The mean change in \MgFe{} from $0-15 \kpc$ is $\approx -.111\dex$ and the host-to-host standard deviation is $\approx 0.02 \dex$.  These differences are likely attributable to our stellar wind model having a metallicity dependent yield for O and not for Mg.  This leads to more O production at small radii where the metallicity is higher, thus a flatter profile.

Fig.~\ref{fig:OonH_z0_sim_v_obs_grad} shows the radial gradients of \OH{} in our simulations at $z = 0$ and includes observations of the \ac{MW}, M31, and nearby \ac{MW}/M31-mass galaxies.
We fit the gradients in our simulations using a least-squares fit of the \OH{} abundance across $4 - 12 \kpc$. As Fig.~\ref{fig:redshift0_radial_summary} shows, including the inner region of our disks, where the bulge dominates ($R \lesssim 4 \kpc$), gives a profile not well approximated by a single linear fit (the bulge is steeper), so we exclude it in fitting this profile, to measure the `disk' component. The range $4 - 12 \kpc$ covers the inner and outer disk and generally exhibits a single power-law profile.
The solid blue line shows the median ($-0.028 \dpk$) across our $11$ galaxies, while the shaded regions show the $68$th percentile and the full distribution. The latter ranges from $-0.042$ to $-0.024 \dpk$.
\FeH{} gradients show similar results, with the full distribution spanning $-0.044$ to $-0.024 \dpk$.

Fig.~\ref{fig:OonH_z0_sim_v_obs_grad} shows \OH{} gradients observed in nearby \ac{MW}/M31-mass galaxies as box-and-whisker plots, with the box showing the $68$th percentile and the whiskers showing the full observed range. We apply a cut on the stellar masses of these observed samples to be comparable to our simulations.
The \citet[][K19]{Kreckel19} sample includes $5$ galaxies from the PHANGS-MUSE survey with $10.2 \leq \log_{10} M_{\rm star} / M_{\odot} \leq 10.6$, the \citet[][Z19]{Zinchenko19} sample includes $7$ galaxies from CALIFA DR3 with $10.2 \leq \log_{10} M_{\rm star} / M_{\odot} \leq 10.8$, the \citet[][B17]{Belfiore17} sample includes $13$ galaxies from the MaNGA survey with $10.2 \leq \log_{10} M_{\rm star} / M_{\odot} \leq 11$, and the \citet[][S16]{SanchezMenguiano16} sample includes $20$ galaxies from the CALIFA survey with $10.2 \leq \log_{10} M_{\rm star} / M_{\odot} \leq 11$.  In addition to the mass cut, we select galaxies that have gradients measured across a radial range comparable to the range in our analysis (the measured ranges all fall within $2 - 14 \kpc$ except for \citet{Kreckel19} which falls within $1 - 11 \kpc$). While all of these observed samples show almost exclusively negative gradients in \OH{}, their abundance gradients are typically flatter than in our simulations. Our simulations are consistent at the $1 \mhyphen \sigma$ level with K19 and B17, and at the $2 \mhyphen \sigma$ level with Z19. However, our sample does not overlap with S16.
Note that the calibrator used for determining the abundances varies from survey to survey.  Using different calibrators can give drastically different abundance measurements \citep{Hemler20}, which could contribute to discrepancies between the different surveys, and to differences with our simulations. Note that the difference between our simulations and these observations are comparable to the differences between surveys themselves.

Fig.~\ref{fig:OonH_z0_sim_v_obs_grad} also shows observed abundance gradients in M31 and the \ac{MW} from HII regions.  The orange points show observed gradients in M31 from \citet[][Z12]{Zurita12} and \citet[][S12]{Sanders12}.  These gradients are slightly shallower than in our simulations, though they agree within $2 \mhyphen \sigma$. 
This may be a consequence of M31 gradient measurements spanning $\approx 4 - 25 \kpc$: from our analysis, including the outer regions of a gas disk flattens the inferred gradient.

The black points show measured gradients of the MW. \citet{Molla19a}, shown in grey, is a best-fit measurement of the \ac{MW} abundance gradient based on the combined data of \citet{Rudolph06, Balser11, Esteban13, Esteban17, Fernandez17}.  We show uncertainties for all samples.  The red error bars for \citet{Balser11, Wenger19} show the impact of measuring the radial gradient along different galactic azimuths.
\citet{Balser11} finds gradients ranging from $-0.03$ to $-0.07 \dpk$ and \citet{Wenger19} find gradients ranging from $-0.035$ to $-0.075 \dpk$ which highlights that measurements of the \ac{MW} radial gradient are strongly sensitive to azimuthal variations.
The different samples include different radial ranges, so they are not exactly comparable to each other or our analysis. Most measurements of \OH{} gradients in the \ac{MW} overlap with our simulations, though our simulations generally have shallower gradients.

While not included in Fig.~\ref{fig:OonH_z0_sim_v_obs_grad}, \citet{Hernandez21} recently measured the radial \OH{} gradient in neutral and ionized gas in M$83$.  The gradients were measured out to $\approx 5.5\kpc$.  They found the gradients in neutral gas to be substantially steeper than the gradients in ionized gas.  As most observations target ionized gas around HII regions, one might expect that our measured gradients shown in Fig.~\ref{fig:OonH_z0_sim_v_obs_grad} are flatter than expected. However, \citet{Hernandez21} measured gradients primarily in the bulge, which we exclude in this analysis. Their gradient for neutral gas external to the bulge is $\approx -0.02 \dpk$ and for ionized gas is $\approx -0.03 \dpk$, in good agreement with our values.

As a whole, the radial gradients in our simulations are somewhat steeper than in external galaxies but somewhat shallower than in the \ac{MW}. The MW may be an outlier: as \citet{Boardman20} note, its gradient is typically steeper than those observed in \ac{MW} analogs.
These differences are likely the result of a combination of different factors, such as: measuring over different radial ranges or using different calibrators.
For example, B17 also measure the gradient in their MaNGA observations using a different calibrator for \OH{} (O$3$N$2$, not shown here, as opposed to R$23$, as  Fig.~\ref{fig:OonH_z0_sim_v_obs_grad} shows), which results in a median gradient that is $\approx 0.008 \dpk$ shallower. Thus, given that S16 used the O$3$N$2$ calibrator applied to their CALIFA observations, this may explain the discrepancy between S16 and B17.
We defer a more detailed comparison via synthetic observations of our simulations, tailored to each observation, to future work.
Rather, Fig.~\ref{fig:OonH_z0_sim_v_obs_grad} provides a broad comparison, highlighting that the radial gradients of gas-phase \OH{} within our simulations lie within the scatter across the \ac{MW}, M31, and nearby \ac{MW}-mass galaxies.

Also, while we do not show it, we compared the \NH{} gradients of our sample at $z = 0$ to observations of M31 and the MW. Our mean \NH{} gradient is $-0.039 \dpk$, with a standard deviation of $0.007$, and the full distribution across hosts spans $-0.028$ to $-0.057 \dpk$.  This agrees well with values measured for the MW in \citet{Esteban18} (comparing 3 different ionization correction factors, they found \NH{} gradients of $-0.047$ to $-0.050 \dpk$ with uncertainties of $\approx 0.008$) and in \citet{Arellano20} ($-0.049 \pm 0.007$). Our \NH{} gradients also agree with the value measured in M31 by \citet{Sanders12} ($-0.0303 \pm 0.0049 \dpk$. However, \NH{} gradients measured in \citet{Rudolph06} ($-0.071 \pm 0.010 \dpk$ in the optical and $-0.085 \pm 0.010 \dpk$ in the far infrared) and \cite{Fernandez17} ($-0.080 \pm 0.019 \dpk$) are substantially steeper than in our hosts. As with the \OH{} gradients, this may be because the gradients are not measured over the same radial range.

\subsection{Evolution of radial gradients}
\label{subsec:radial_gradient_evolution}

We next explore the evolution of gas-phase radial gradients of \OH, \FeH, and \NH{} at $z \leq 1.5$, over the last $\sim 10 \Gyr$, during the primary epoch of disk assembly, to understand the initial conditions for star formation and chemical tagging of stars.
In summary, we find that at earlier times the gas disk was more radially homogeneous (flatter gradients), so chemical tagging offers less discriminating power for radial birth location at earlier times.

\begin{figure}
    \includegraphics[width = 0.91 \columnwidth]{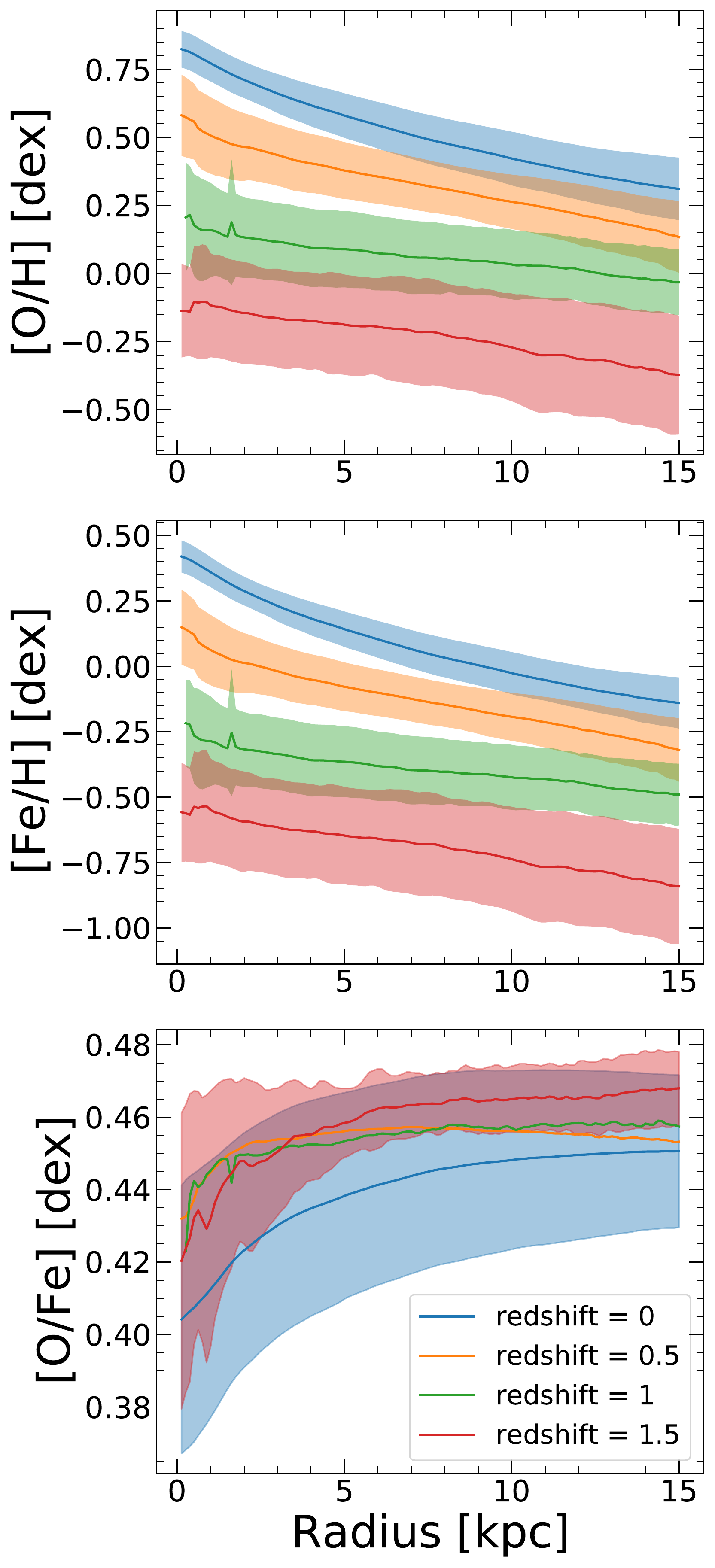}
    \vspace{-3 mm}
    \caption{
    Radial profiles of gas-phase elemental abundances in our simulations at different redshifts. The solid lines show the mean and the shaded regions show the $1 \mhyphen \sigma$ scatter across our $11$ galaxies. The top and middle panels show steepening radial profiles of \OH{} and \FeH{} with time, as the gas disk becomes more rotation dominated, with less radial turbulence, thus sustaining stronger radial gradients. The bottom panel shows that the innermost regions of the gas disk have lower \OFe{} than the outer disk, this indicates that the inner disk is more evolved, that is, has had more Ia supernovae that produce more Fe than $\alpha$-elements (like O), than the outer disk.
    }
    \label{fig:radial_mf_summary}
\end{figure}

Similar to Fig.~\ref{fig:redshift0_radial_summary}, Fig.~\ref{fig:radial_mf_summary} shows radial profiles of \OH, \FeH, and \OFe{} in gas at different redshifts.
The solid line shows the mean across our 11 galaxies, while the shaded region shows the $1 \mhyphen \sigma$ scatter.
At all radii, \OH{} and \FeH{} increase with time, as the gas mass declines while more stars enrich the \ac{ISM}. This evolution agrees with the observed gas-phase galaxy mass-metallicity relation \citep{Tremonti04}.
\citet{Ma16} explored this evolution across a wide galaxy mass range in the \ac{FIRE}-1 simulations: they found that as galaxies grow more massive, the mass-loading factor of their winds decreases, and metals are more easily held in/near the galaxy as opposed to being driven into the halo \citep[see also][]{Muratov15, Muratov17, AA17}.

Fig.~\ref{fig:radial_mf_summary} also shows that both \OH{} and \FeH{} have negative radial gradients at all times. \citet{Ma17} also found primarily negative gradients in the \ac{FIRE}-1 suite, because the high star-formation efficiency in the inner disks of galaxies with well ordered rotation leads to sustained negative radial gradients.
At $z = 0$, our average change in \OH{} from $0 - 15 \kpc$ is $\approx 0.51 \dex$, while this is $\approx 0.56 \dex$ for \FeH{}.
At $z = 1.5$ ($t_{\rm lookback} = 9.4 \Gyr$), the average change in abundance from $0 - 15\kpc$ is $\approx 0.24 \dex$ for \OH{} and $0.28 \dex$ for \FeH{}.
Furthermore, as expected given the scatter in formation history, we find larger host-to-host scatter (averaged over all radii) at earlier times: $0.09 \dex$ for \OH{} and $0.07 \dex$ for \FeH{} at $z = 0$, but at $z = 1.5$ this was $0.2 \dex$ for both elements.

\begin{figure}
	\includegraphics[width = \columnwidth]{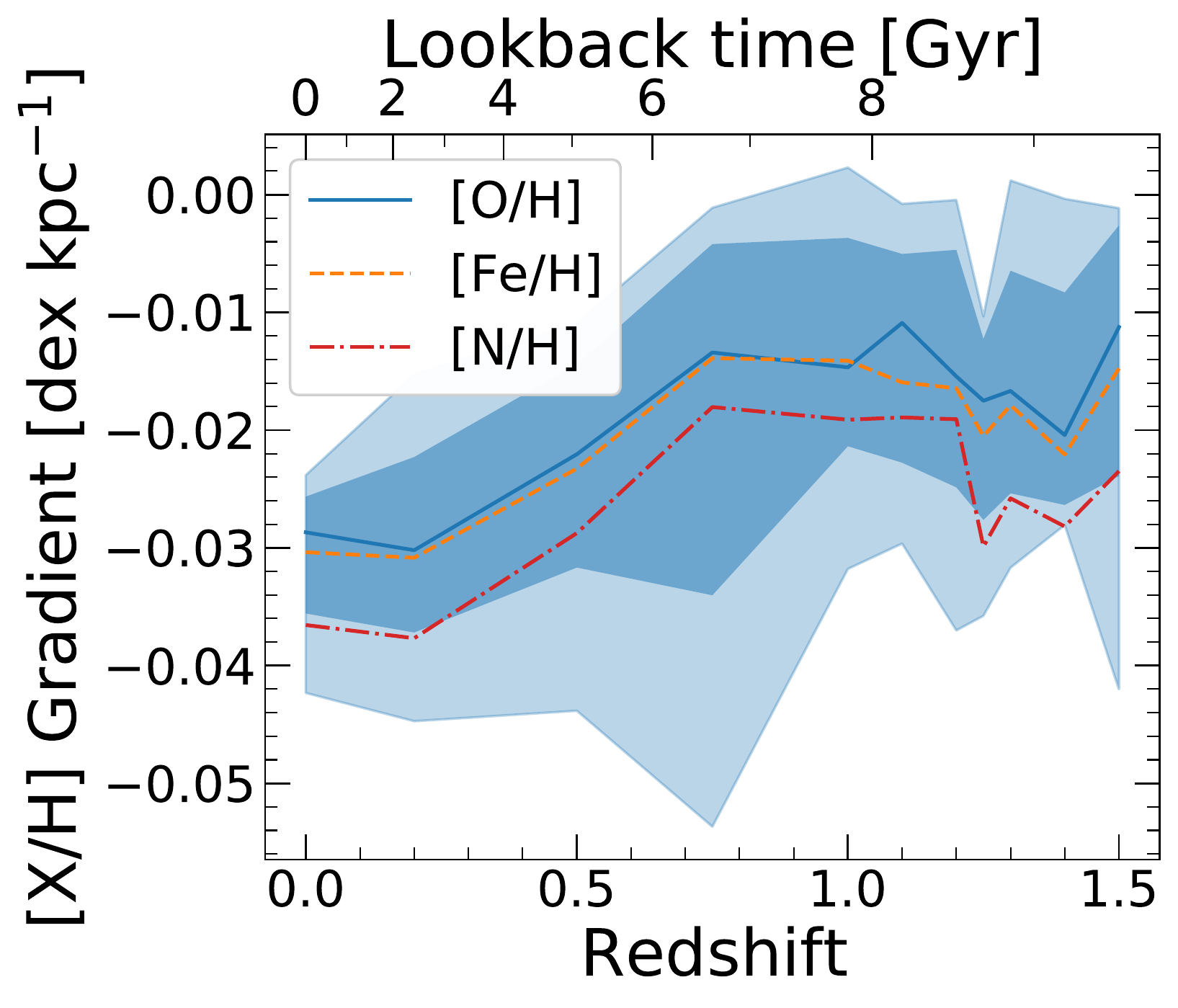}
	\vspace{-6 mm}
    \caption{
    Evolution of radial gradients in gas elemental abundances. The lines show the means for \OH, \FeH, and \NH.  The dark shaded region shows the $1 \mhyphen \sigma$ scatter and the light shaded region shows the total scatter in \OH{} across our 11 galaxies. At each redshift we calculate the gradient via a linear fit across a redshift-dependent radial range: $4 < R < 12 \kpc$ for $z < 1$ ($t_{\rm lookback} < 7.8 \Gyr$) and $0 < R < 8 \kpc$ for $z \geq 1$. The gradient is flattest ($\approx -0.015 \dpk$) at $z = 1.5$ ($t_{\rm lookback} = 9.4 \Gyr$), likely because high merger rates drive large scale radial turbulence, erasing the radial gradients of the disks.  This leads to the azimuthally averaged metallicity at a given radius showing little radial dependence although stochastic enrichment/accretion events may lead to locally over or under-enriched regions. Shallow gradients persist until $z \approx 1$.  At lower redshifts, the gas disk becomes more rotationally supported and is capable of sustaining stronger radial gradients, becoming steepest ($\approx -0.03 \dpk$) at $z = 0$. The gradients of \OH{} and \FeH{} are almost identical, despite being sourced primarily by core-collapse and Ia supernovae, respectively. However, the \NH{} gradient, sourced primarily through stellar winds from massive stars, which have a metallicity dependent mass loss rate, is steeper at all times.
    }
    \label{fig:radial_grad_evolution}
\end{figure}

Fig.~\ref{fig:radial_mf_summary} also shows that the abundance profiles were flatter (more homogeneous) at earlier times, because the abundance at smaller radii evolves more rapidly than at larger radii.
Increased accretion/merger rates, coupled with higher star-formation rates and stronger gas turbulence, drove more efficient radial mixing at earlier times \citep{Ma17}. In FIRE simulations, early galaxies experience bursty, stellar feedback-driven outflows that radially mix the ISM, in addition to local turbulence. The profiles steepen with time, because as the gas disk settles down and becomes more rotationally supported, it is capable of sustaining stronger radial gradients given less radial mixing \citep{Ma17}. At $z \lesssim 1$ ($t_{\rm lookback} \lesssim 7.8 \Gyr$), the radial profiles generally show an up-turn in the innermost bulge-dominated region. Beyond $\approx 4 \kpc$, the profiles are well fit by a linear relation at all redshifts.

Fig.~\ref{fig:radial_mf_summary} (bottom panel) shows that \OFe{} tends to decline over time at all radii, because at early times, core-collapse supernovae dominate the enrichment, which preferentially produce $\alpha$-elements like O.  
However, the mean trends in Fig.~\ref{fig:radial_mf_summary} (bottom panel) are not necessarily true for individual hosts.  The dynamic range of \OFe{} is small, so minor upticks in star formation rates lead to enhanced \OFe{} as the rate of core collapse supernovae temporarily increases. Individual galaxies show overall increases in \OFe{} following periods of increased star formation, and different radii of individual hosts show relative increases or decreases in \OFe{}, likely correlated with radial variations in star-formation rates.

On average, at later times, the (more delayed) Ia supernovae preferentially enrich the galaxy in Fe, driving down \OFe{}.
However, the typical change in gas-phase \OFe{} at fixed radius from $z = 1.5$ ($t_{\rm lookback} = 9.4 \Gyr$) to 0 is only $\approx 0.02 - 0.03 \dex$.
The \OFe{} radial gradients are positive at all times, because the outer disk is always younger than the inner disk/bulge, though the gradients are weak at larger radii.
The \OFe{} radial profiles steepen at small radii at later times, at least at $z \lesssim 1$ ($t_{\rm lookback} \lesssim 7.8 \Gyr$).
Unlike the profiles of individual elements, the host-to-host standard deviation of \OFe{} increases at later times, from $0.015 \dex$ at $z = 1.5$ to $0.037 \dex$ at $z = 0$.
Overall, \OFe{} does not provide strong discrimination power for chemical tagging at any radii or time that we examine.

While not shown here, we also measure the evolution of \MgFe{}, which is more significant than \OFe{}. In the outer disk ($R = 12 \kpc$) \MgFe{} decreases from $\approx 0.3 \dex$ at $z = 1.5$ to $\approx 0.22 \dex$ at $z = 0$. In the inner disk ($R = 4 \kpc$) the evolution is larger, from $\approx 0.29 \dex$ to $0.18 \dex$ over the same redshifts. The stronger evolution seen in \MgFe{} likely results from stellar winds in our simulations producing relatively little Mg. The stellar-wind model in FIRE-2 has a metallicity-dependent yield for O. At lower redshifts, as the gas disk becomes more enriched, O production also increases, leading to less evolution in \OFe{}.

Fig.~\ref{fig:radial_grad_evolution} shows the evolution of the mean radial gradients of \OH{}, \FeH{}, and \NH{} at redshifts $0$, $0.2$, $0.5$, $0.75$, $1$, $1.1$, $1.2$, $1.25$, $1.3$, $1.4$, and $1.5$. The shaded regions show the $1 \mhyphen \sigma$ scatter. We also checked that the evolution of each galaxy qualitatively agrees with this mean evolution. We measure radial gradients between $4 < R < 12 \kpc$ at $z < 1$ ($t_{\rm lookback} < 7.8 \Gyr$), consistent with Fig.~\ref{fig:OonH_z0_sim_v_obs_grad}, and between $0 < R < 8 \kpc$ at $z \geq 1$ ($t_{\rm lookback} \geq 9.4 \Gyr$). (For a few hosts with bulge-like upturns in their profiles at $\lesssim 2 \kpc$, we measure their gradients between $2 - 8 \kpc$, where the gradient is nearly linear).
We use a redshift-dependent radial range, because we are exploring the gas from which stars are forming. We select gas approximately contained in $R^{\rm star}_{90}$, which is $\approx 11 \kpc$ at $z = 0$ and $\lesssim 8 \kpc$ at $z \gtrsim 0.5$. At $z < 1$, we measure the gradient at $4 - 12 \kpc$, because the inner disk is dominated by the bulge region and has a steeper profile, as Fig.~\ref{fig:radial_mf_summary} shows.
Thus, at late times we measure at $4 - 12 \kpc$, which encompasses both the inner and outer disk and exhibits a nearly linear profile.
We tested our analysis measuring the gradient over varying radial ranges and found that, while the normalization varies somewhat, the shape of the profile and the evolution are consistent, as Fig.~\ref{fig:radial_mf_summary}.

As Fig.~\ref{fig:radial_grad_evolution} shows, the strength of the radial gradient generally decreases over time.
The minimum magnitude of the gradient is $\approx 0.01 \dpk$ and occurs at $z \approx 1.5$ ($t_{\rm lookback} \approx 9.4 \Gyr$). We find just 2 galaxies that achieve a flat gradient at this time. At $z \lesssim 1$ ($t_{\rm lookback} \lesssim 7.8 \Gyr$), the radial gradients gradually steepen to $-0.03 \dpk$ at $z = 0$. The gradients prior to $z \approx 1$ are approximately constant with redshift.  While there are fluctuations in the gradient at high $z$, these are transient features in the simulations driven primarily through minor mergers, flybys of satellite galaxies, and starbursts.
The host-to-host scatter is smallest at $z = 0$ and is largest at $z = 0.75$ ($t_{\rm lookback} = 6.6 \Gyr$), in part because of one galaxy (m12f) that experiences a major merger at this time.

Fig.~\ref{fig:radial_grad_evolution} also compares the evolution of the radial gradients in \OH{} and \FeH{} with \NH.
Consistent with most results in this paper, we find little-to-no difference between \OH{} and \FeH{}, despite their differing origins, from primarily core-collapse and primarily Ia supernovae, respectively.
However, we find systematically stronger radial gradients in \NH{} at all times.
Unlike O and Fe, which are sourced primarily through supernovae, N is sourced primarily by stellar winds in the FIRE-2 simulations, and 
the wind mass-loss rate from massive stars (in the first $3.5 \Myr$) depends roughly linearly on progenitor metallicity.
(The N yield from core-collapse supernovae also increases linearly with progenitor metallicity in the FIRE-2 model, but this effect is subdominant, because most N comes from stellar winds.)
The progenitor metallicity dependence of N (often called secondary production of N) results in enhanced N production in regions that are already more metal rich, and thus it drives a steeper gradient for N (by $\approx 0.015 \dpk$) than O or Fe at all times.

\subsection{Vertical profiles across time}

\begin{figure}
	\includegraphics[width=\columnwidth]{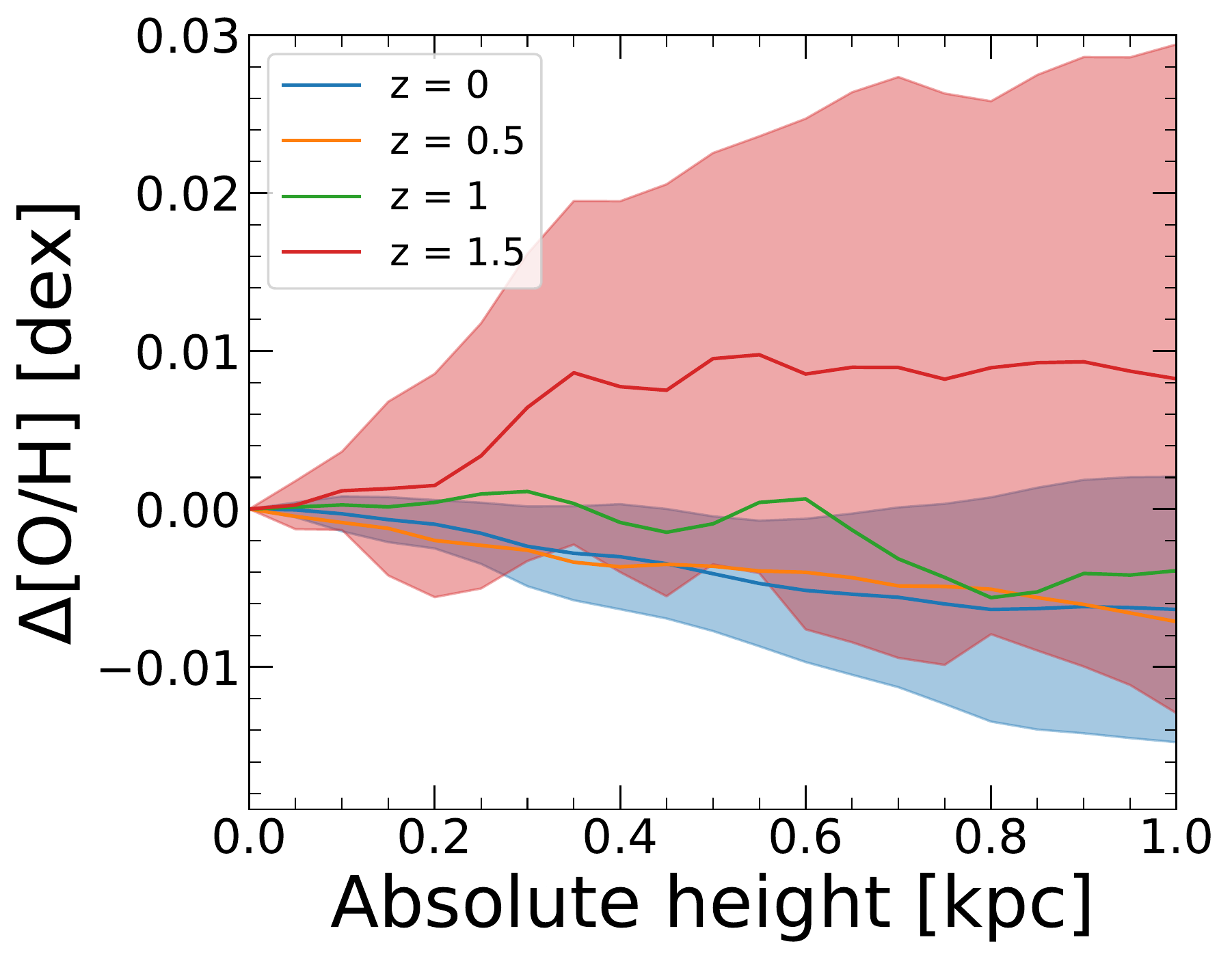}
	\vspace{-6 mm}
    \caption{
    The vertical change in \OH{} relative to the midplane at various redshifts.
    We measure the vertical profile at $R = 7 - 9 \kpc$. The solid lines show the mean while the shaded region shows the $1 \mhyphen \sigma$ standard deviation across 11 galaxies. \OH{} shows minimal variation with height: the mean deviation at $1 \kpc$ above the midplane is $\approx 0.008 \dex$ at $z = 1.5$ ($t_{\rm lookback} = 9.4 \Gyr$) and $\approx -0.006 \dex$ at $z = 0$. The majority of stars form within a few hundred parsecs of the galactic midplane at low $z$ and within $\sim 1.5 \kpc$ of the midplane at the highest redshift we examine.  Close to the plane of the disk ($|Z| \leq 200 \pc$) the variation in \OH{} is even smaller at all  times ($|\Delta \rm [O/H]| \lesssim 0.001 \dex$ at $z = 1.5$ and $|\Delta \rm [O/H]| \lesssim 0.002 \dex$ at $z = 0$). At $z < 1$ ($t_{\rm lookback} < 7.8 \Gyr$) we find a weak but systematically negative gradient.
    We conclude that vertical gradients are not measurably significant for chemical tagging at any redshift.
    }
    \label{fig:vertical_evolution}
\end{figure}

We next examine the vertical profiles (in absolute height) of elemental abundances, for all gas near the solar circle, within a cylindrical radius of $7 < R < 9 \kpc$.
We normalize the vertical profiles by subtracting the midplane abundance at each redshift. Fig.~\ref{fig:vertical_evolution} shows the vertical profiles for \OH{}. The solid line shows the mean and the shaded regions show the $1 \mhyphen \sigma$ scatter; we only show the scatter at $z = 1.5$ ($t_{\rm lookback} = 9.4 \Gyr$) and $z = 0$ for clarity.

Fig.~\ref{fig:vertical_evolution} shows that any systematic trends in abundance with height to $1 \kpc$ is $\lesssim 0.01 \dpk$ absolute on average at all times, and the $1 \mhyphen \sigma$ scatter is $\lesssim 0.01 \dpk$ at $z = 0$ and $\lesssim 0.02 \dpk$ at $z = 1.5$. Thus, the gas disk is well mixed vertically.
In most of our galaxies, the deviations in abundance increase with distance from the midplane, that is, they shows a systematic gradient with height.
Over time, these vertical gradients become increasingly (if weakly) more negative, which supports the idea of `upside-down' disk formation \citep[e.g.][]{Bird13, Ma17, Bird21}, such that stars formed in a more vertically extended disk at higher redshifts, leading to more enrichment at larger heights, at later times stars formed in a thinner disk and gas farther out of the midplane became relatively less enriched.
At $z \gtrsim 1$, the absolute strength of this vertical gradient is in fact comparable to the radial gradient (Fig.~\ref{fig:radial_grad_evolution}), while at $z = 0$, the vertical gradient is $\approx 3 \times$ weaker than the radial gradient.
This is because the timescale for vertical mixing is short, given gas turbulence, and that the vertical scale-height of the gas is itself set by the maximum Jeans length at that time.
Furthermore, with implications for chemical tagging, the majority of star formation in our simulations is limited to $\lesssim 500 \pc$ of the midplane at $z < 0.5$, and $\lesssim 1.5 \kpc$ up to $z < 1.5$, and Fig.~\ref{fig:vertical_evolution} shows that vertical variations in abundance are minimal on those scales.
The vertical trends in \FeH{} (not shown here) are consistent with \OH{} within $\approx 0.01 \dex$.

\subsection{Azimuthal variations across time}

\begin{figure*}
    \includegraphics[width=0.95\textwidth]{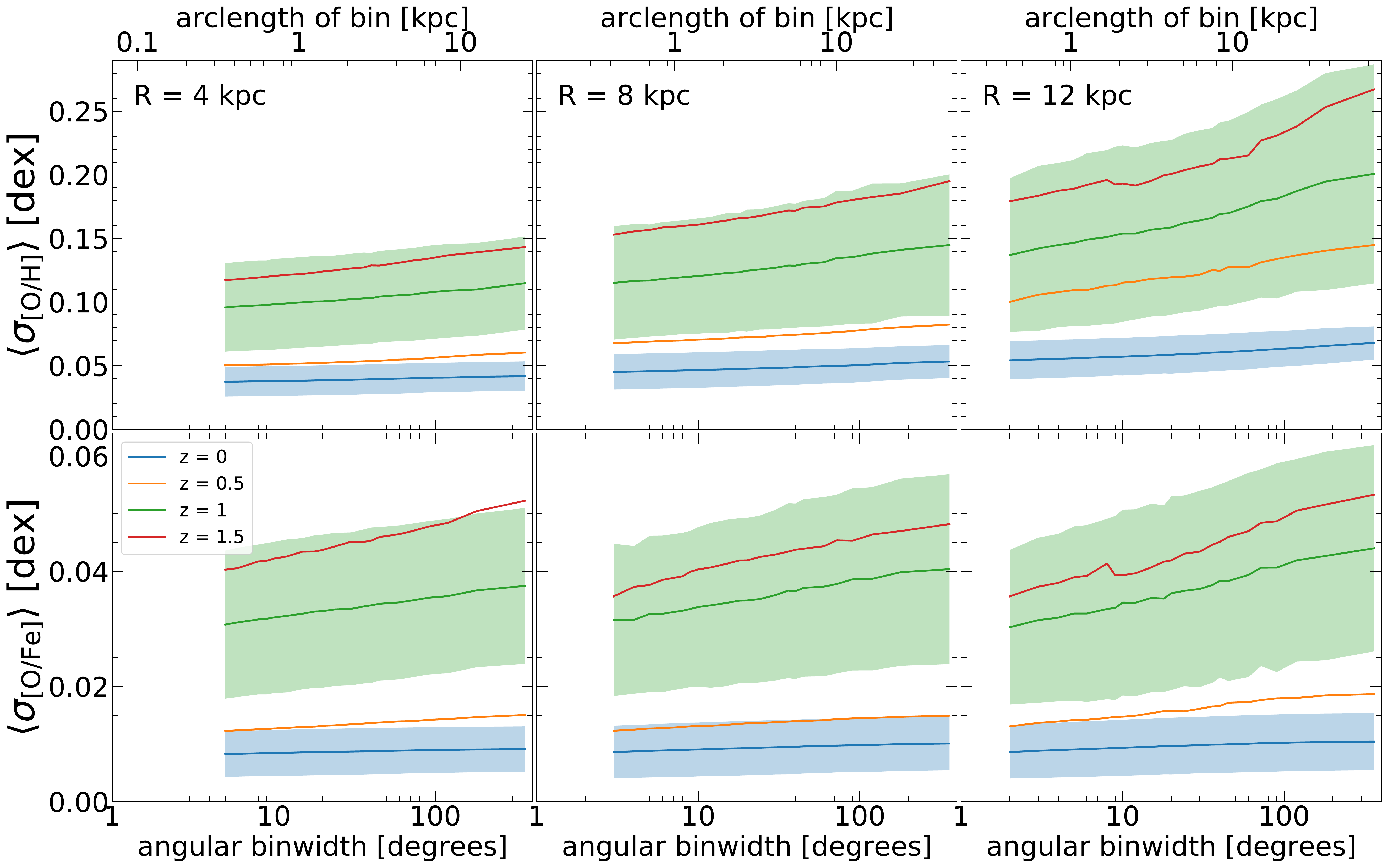}
    \vspace{-2 mm}
    \caption{
    Azimuthal scatter in elemental abundances for \OH{} and \OFe{} in gas, as a function of angular scale, at different redshifts and different radii.
    The solid lines show the mean and the shaded regions show the $1 \mhyphen \sigma$ scatter across our suite of 11 galaxies: we show scatter only at $z = 1$ ($t_{\rm lookback} = 7.8 \Gyr$) and 0. The scatter increases as a function of angular bin size at all redshifts and at all radii. At $z = 0$, near the solar circle ($R = 8 \kpc$), the average azimuthal scatter across the disk is $\approx 0.053 \dex$ for \OH{} and \FeH{} (not shown) and $\approx 0.009 \dex$ for \OFe. For all angular bin sizes, the average scatter increases with redshift: at earlier cosmic times the gas disks were less well mixed within a given annulus. At $z = 0$ the scatter across the disk is $\approx 0.2 \dex$ for \OH{} and \FeH{} (not shown) and $\approx 0.05 \dex$ for \OFe. The scatter also increases with angular bin size at all redshifts, although the increase is minimal at late times. At low $z$, this means that azimuthal variations are dominated by local (and not global) fluctuations in the disk. Finally, the azimuthal scatter increases with radius for individual abundances: gas is azimuthally better mixed in the inner disk, likely a result of shorter orbital times leading to faster mixing.
    }
    \label{fig:azimuthal_scatter_summary}
\end{figure*}

\begin{figure}
	\includegraphics[width=\columnwidth]{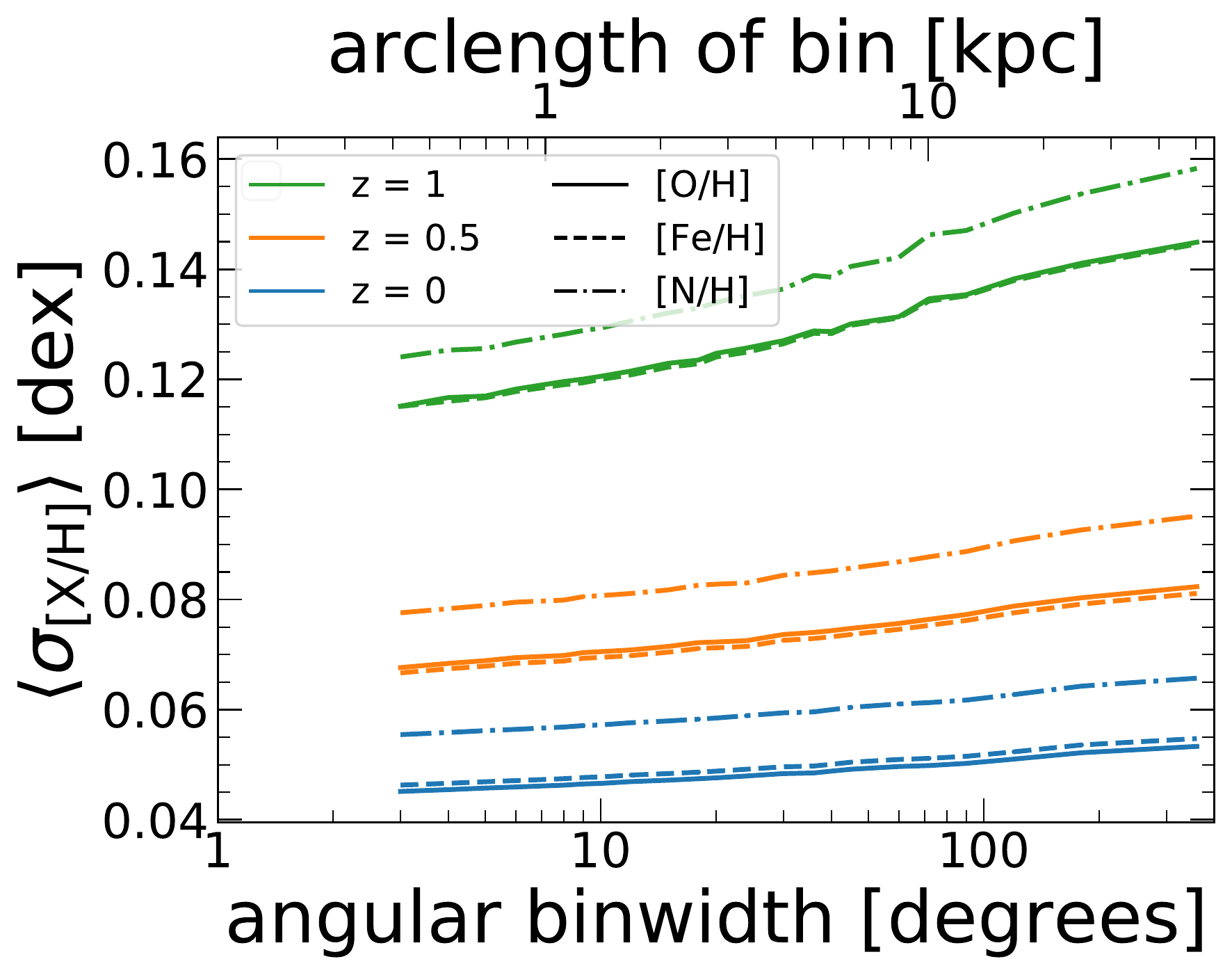}
	\vspace{-6 mm}
    \caption{
    Azimuthal scatter, as in Fig.~\ref{fig:azimuthal_scatter_summary}, at $R = 8 \kpc$ for \OH{} (solid), \FeH{} (dashed), and \NH{} (dash dot). 
    As with the radial gradients, the azimuthal variations of \OH{} and \FeH{} are almost identical, despite being sourced primarily by core-collapse and Ia supernovae, respectively, while the variation in \NH{}, which is sourced primarily through metallicity-dependent stellar winds from massive stars, is higher at all redshifts.
    }
    \label{fig:azimuthal_scatter_NonH_comparison}
\end{figure}

We next investigate azimuthal variations of elemental abundances in gas, including its evolution. We thus test a common assumption in galactic evolution, that gas is well mixed azimuthally within a given annulus \citep[e.g.][]{Frankel18, Frankel20}.

Fig.~\ref{fig:azimuthal_scatter_summary} shows the standard deviation in \OH{} and \OFe{} along angular bins of varying length at fixed radius.
Specifically, we compute the standard deviation within a given angular bin, and Fig.~\ref{fig:azimuthal_scatter_summary} shows the mean standard deviation across all bins of a given angular size for all 11 simulations. We stack 3 snapshots ($\Delta t \approx 50 \Myr$) for each redshift.
We use an annulus of gas $\pm 0.3 \kpc$ out of the plane of the disk
because as shown in Fig.~\ref{fig:vertical_evolution} gas-phase abundances are effectively homogeneous within this height.
We also measure within a radius $\pm 0.15 \kpc$ of the selected cylindrical radius, to ensure that the angular length dominates over the radial length for our smallest angular bins, that is, to ensure that the radial gradient does not induce significant scatter.
We show the $1 \mhyphen \sigma$ scatter for $z = 0$ and $z = 1$ ($t_{\rm lookback} = 7.8 \Gyr$).
We exclude m12c at $z = 1.5$ for angular scales $\Delta \phi \leq 8^{\circ}$, because its angular bins contain too few gas particles.

At $z = 0$ and $R = 8 \kpc$ (near the solar circle) the typical azimuthal scatter across the gas disk is $\lesssim 0.053 \dex$ for \OH{}, $\lesssim 0.055 \dex$ for \FeH{} (not shown), and $\lesssim 0.01 \dex$ for \OFe. This value for \OH{} agrees well with \ac{MW} observations \citep{Wenger19} and observations of external galaxies \citep{Sakhibov18, Kreckel19, Kreckel20}, though we emphasize that we are not measuring azimuthal scatter in the same way: those observations typically measure differences in abundances between arm and inter-arm regions or measure abundance variations between HII regions within an aperture of a given size.

Our azimuthal scatter decreases with smaller angular bin size, with a minimum of $\approx 0.045 \dex$ for \OH{} ($\approx 0.046 \dex$ for \FeH{}) and $\approx 0.009 \dex$ for \OFe at the smallest angular scales. Interestingly, this minimal scatter remains well above $0 \dex$ as $\Delta \phi$ goes to $0$. We emphasize that our analysis does not zoom-in on \ac{GMC} or individual star-forming regions, but rather we examine all of the \ac{ISM} centered on (effectively) random positions. Thus, our results on small scales do not immediately inform the homogeneity of individual \ac{GMC}s, especially given their short lifetimes ($\lesssim 7 \Myr$) in our simulations \citep{Benincasa20}, and we will examine \ac{GMC} homogeneity in future work.
Appendix.~\ref{subsec:all_v_star_forming_gas} also examines how small-scale variations depend on our choice of diffusion coefficient for sub-grid turbulent mixing in gas.

The $1 \mhyphen \sigma$ host-to-host scatter is approximately independent of bin size and is $\lesssim 0.014 \dex$ for \OH, $\lesssim 0.015 \dex$ for \FeH, and $\lesssim 0.005 \dex$ for \OFe. Thus, at $z = 0$ gas within all of our galaxies is well mixed, that is, the azimuthal scatter is comparable to typical measurement uncertainties ($\sim 0.05 \dex$) for elemental abundances.

Fig.~\ref{fig:azimuthal_scatter_summary} shows that, at all radii and at all angular bin sizes, the azimuthal scatter was more significant at earlier times, that is, gas was less azimuthally mixed than it is now.
This is likely because higher accretion and star formation rates combined with burstier star formation leads to more pronounced local pockets of enrichment in gas. At $R = 8 \kpc$ and at $z = 1.5$ ($t_{\rm lookback} = 9.4 \Gyr$) the azimuthal scatter across the disk is $\lesssim 0.2 \dex$ for \OH and \FeH{} (not shown) and $\lesssim 0.05 \dex$ for \OFe.  The scatter does not drop below $\approx 0.15 \dex$ for either \OH{} or \FeH{} at the smallest azimuthal scales ($0.035 \dex$ for \OFe).  The $1 \mhyphen \sigma$ host-to-host scatter is $\lesssim 0.05\dex$ for \OH{} and \FeH{} and $\lesssim 0.016 \dex$ for \OFe.

Additionally, Fig.~\ref{fig:azimuthal_scatter_summary} shows that the difference in scatter between large and small angular scales varies with time. This difference is more significant at earlier times: at $z \gtrsim 1$ ($t_{\rm lookback} \gtrsim 7.8 \Gyr$) this change is $\approx 0.042 \dex$ at $R = 8 \kpc$. Thus, at early times, galaxy-scale fluctuations are more important in driving azimuthal scatter (as visible in Fig.~\ref{fig:gas_map}).
However, at $z \sim 0$, the azimuthal scatter across small versus large angular scales differs by only $\approx 0.008 \dex$, so small-scale fluctuations drive most of the azimuthal scatter (also visible in Fig.~\ref{fig:gas_map}).
These results at low redshift are useful from an observational perspective, because they means that one can generalize smaller-scale observations of gas-phase abundances to overall azimuthal trends at fixed radius.

Fig.~\ref{fig:azimuthal_scatter_summary} also shows that the azimuthal scatter depends on radius. The azimuthal scatter increases with increasing radius for a given angular bin size, and in fact, this is true for both fixed angular and physical bin size. At $z = 0$ the azimuthal scatter across the entire disk at $R = 4 \kpc$ is $\lesssim 0.042 \dex$ for \OH{} ($\lesssim 0.046 \dex$ for \FeH{}, not shown), and this increases to $\lesssim 0.062 \dex$ for \OH{} and \FeH{} at $R = 12 \kpc$.  At $z = 1.5$ ($t_{\rm lookback} = 9.4 \Gyr$) the azimuthal scatter ranges from $\lesssim 0.015 \dex$ to $0.25 \dex$ for \OH{} and \FeH{}. We do not find radial dependence in \OFe{}, which has a maximal scatter $\lesssim 0.01 \dex$ ($\lesssim 0.052 \dex$) at all radii at $z = 0$ ($z = 1.5$).  Most likely, the radial dependence in the azimuthal scatter of \OH{} and \FeH{} results from the increase of the orbital timescale, and hence the timescale for mixing, with radius. Furthermore, cosmic accretion of under-enriched gas also likely contributes to this increase in azimuthal scatter with radius, especially at earlier times, when the increase with radius is stronger.

Fig.~\ref{fig:azimuthal_scatter_NonH_comparison} shows the azimuthal scatter of \OH, \FeH, and \NH{} at $R = 8 \kpc$ at 3 redshifts. As we discussed above, metallicity-dependent stellar winds from massive stars, rather than supernovae, primarily source N, so this compares the azimuthal mixing of elements sourced by these different processes. The azimuthal scatter of \NH{} is larger than that of \OH{} and \FeH{} for all bins at each redshift. On the scale of the entire annulus, the scatter in \NH{} is approximately $0.015 \dex$ larger at $z = 1$ ($t_{\rm lookback} = 7.8 \Gyr$) and approximately $0.01 \dex$ larger at $z = 0$.  This discrepancy is slightly smaller for smaller angular scales at $z = 1$ (approximately $0.01 \dex$ difference), but the difference in azimuthal scatter is independent of scale at $z = 0$. As previously stated, our stellar-wind rate depends linearly on progenitor metallicity, which likely drives these differences for N as compared with O and Fe. One might expect Fe, being sourced primarily by Ia supernovae, to have less scatter at small scales than O, because Ia are caused by preferentially older stars than core-collapse supernovae, which occur closer to stellar birth location. A possible cause of our similarity comes from our assumed Ia delay-time distribution \citep{Mannucci06}, which has a significant component from prompt Ia, at ages $\lesssim 100 \Myr$.
A Ia delay-time distribution with a more significant contribution from older stellar ages \citep[][]{Maoz17} may lead to less small-scale scatter (e.g. Gandhi et al., in prep.). 

\subsection{Azimuthal versus radial variations across time}
\label{subsec:azimuthal_vs_radial}

\begin{figure*}
    \begin{minipage}{.48\textwidth}
        \centering
        \includegraphics[width = \columnwidth]{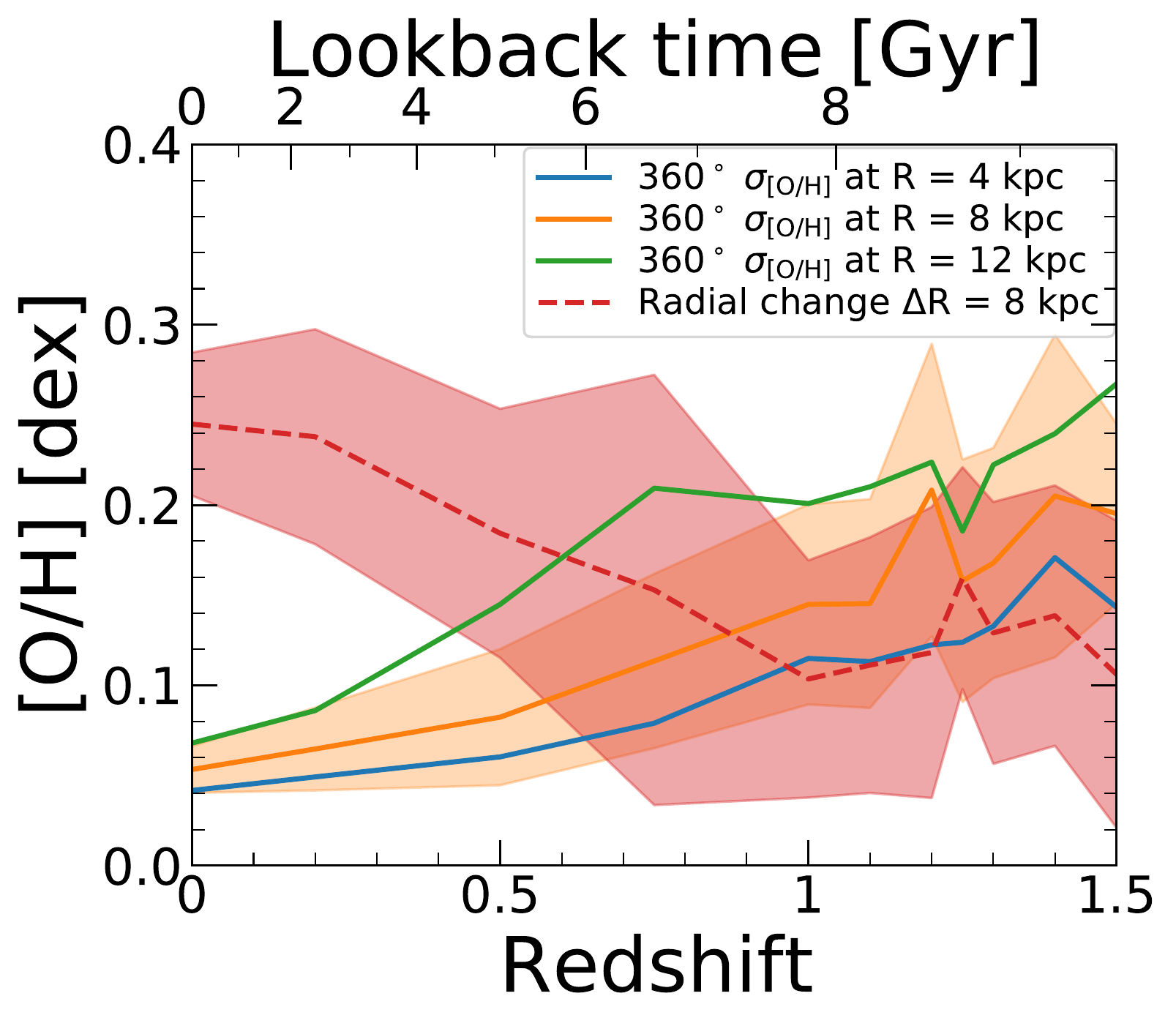}
        \vspace{-6 mm}
        \caption{
        The evolution of variations in \OH{} in gas, both radially and azimuthally. (We do not show \FeH{}, its trends are consistent with \OH{} to $\lesssim 0.01 \dex$.) The solid lines show the mean scatter, across our 11 galaxies, for the full ($360^\circ$) annulus of gas at $R = 4 \kpc$ (blue), $R = 8 \kpc$ (yellow), and $R = 12 \kpc$ (green). The red dashed line shows the mean radial change in \OH{} across a radial distance of $8 \kpc$. The shaded regions show the $1 \mhyphen \sigma$ scatter. While the radial gradient dominates the spatial variations at late cosmic times,  azimuthal variations were more significant than the radial gradient at earlier cosmic times ($z \gtrsim 0.9$, or $t_{\rm lookback} \gtrsim 7.4 \Gyr$, at $R = 8 \kpc$).
        Larger radii transition to radially dominated abundance variations at later times (see also Fig.~\ref{fig:z_grad_dominated}).
        Thus, elemental evolution models should not assume azimuthal homogeneity of abundances at early times; instead, azimuthal variations are the primary source of spatial information for chemical tagging of stars forming at early times.
        }
        \label{fig:radial_v_azimuthal}
    \end{minipage}
    \hfill
    \begin{minipage}{0.48\textwidth}
        \centering
        \includegraphics[width=\columnwidth]{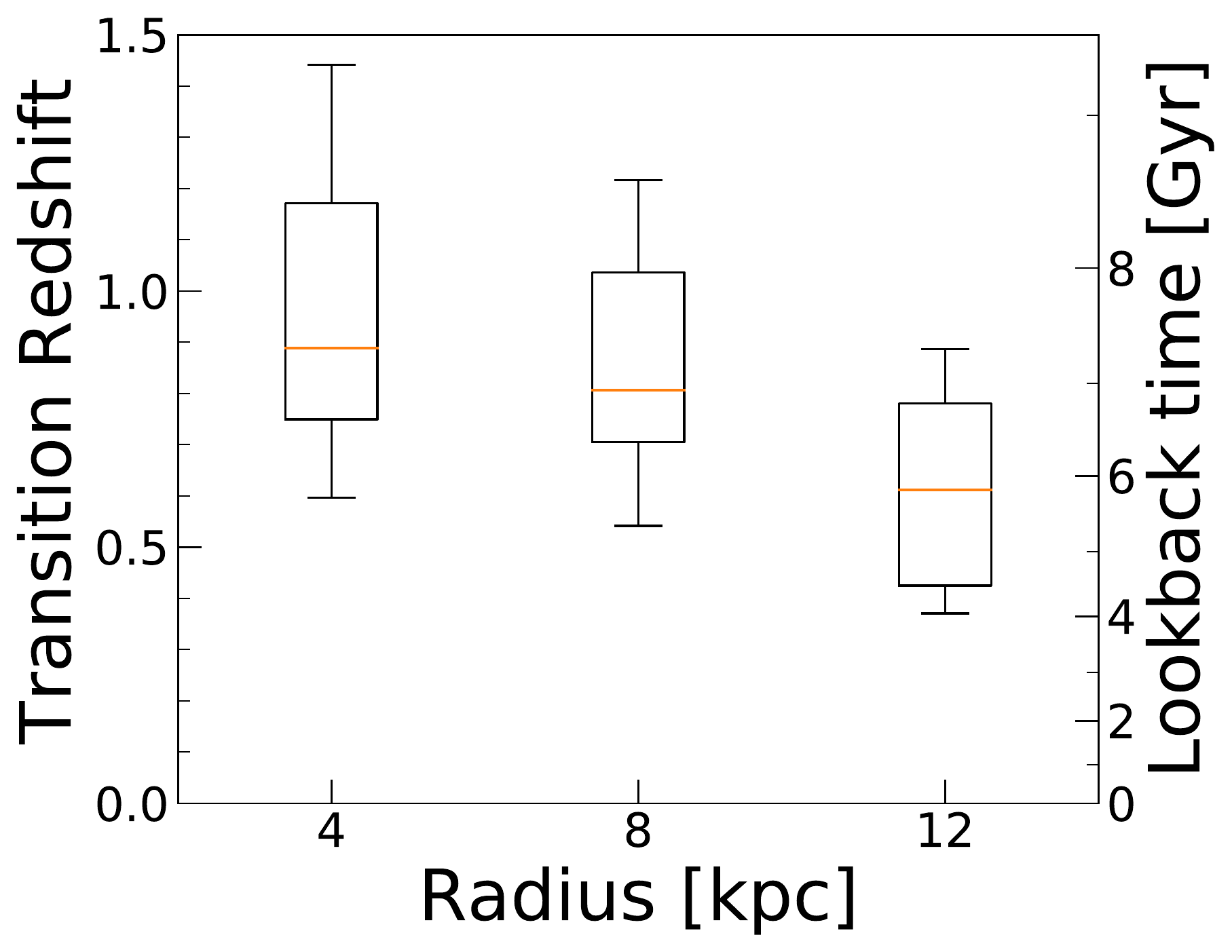}
        \vspace{-6 mm}
        \caption{
        Following Fig.~\ref{fig:radial_v_azimuthal}, the redshift below which radial variations in \OH{} dominate over azimuthal variations at $3$ radii in our $11$ simulations (the intersection of the dashed and solid lines). The horizontal line shows the median, the box shows the $68$th percentile, and the whiskers show the full distribution.
        This transition redshift is the last time the azimuthal variation is stronger than the radial variation.  This transition occurs earlier at smaller radii, where azimuthal variations are smaller (Fig.~\ref{fig:azimuthal_scatter_summary}), given the shorter timescale for mixing at smaller radii. Before these transition redshifts, any model of chemical tagging should account for azimuthal variations as the primary source of spatial variation.
        }
        \label{fig:z_grad_dominated}
    \end{minipage}
\end{figure*}

We now compare the relative importance of radial gradients versus azimuthal scatter in gas-phase abundances.
Fig.~\ref{fig:radial_v_azimuthal} shows the evolution of the radial variations in \OH{}, from multiplying the radial gradient (as calculated in \S\ref{subsec:radial_gradient_evolution}) of each simulation at each redshift by $8 \kpc$, to show the change from the inner to the outer disk. This measures the change from $4 - 12 \kpc$ at $z < 1$ ($t_{\rm lookback} < 7.8 \Gyr$ and from $0 - 8 \kpc$ at $z \geq 1$, where the radial profile is approximately linear (see Fig.~\ref{fig:radial_mf_summary}).
Fig.~\ref{fig:radial_v_azimuthal} also shows evolution of the azimuthal scatter around each disk ($360^\circ$) at 3 radii.

Fig.~\ref{fig:radial_v_azimuthal} shows that at early times, $z \gtrsim 1$ ($t_{\rm lookback} \gtrsim 7.8 \Gyr$), the $360^\circ$ angular scatter in \OH{} at all radii is larger than the radial variation, with the angular scatter at large radii being approximately $2 \times$ higher. However, the radial variation dominates at all radii at $z \lesssim 0.6$ ($t_{\rm lookback} \lesssim 5.8 \Gyr$).  At $z = 0$ the radial variations in \OH{} is approximately a factor of 4 higher than the azimuthal variation. 
Thus, at $z \sim 0$, one can approximate the gas disk variations primarily via the radial gradient, but at earlier times, the azimuthal variations dominate.

Fig.~\ref{fig:z_grad_dominated} shows the redshift (and lookback time) when the radial variation begins to dominate over the azimuthal scatter in \OH{} (the intersection of the solid and dashed lines in Fig.~\ref{fig:radial_v_azimuthal}), for 3 radii.
We measure this transition redshift separately for each simulation: the horizontal lines show the median, and the boxes and whiskers show the $68$th percentile and the full distribution. 
The radial variation starts to dominate earlier at smaller radii, because the azimuthal scatter increases with radius at all redshifts (Fig.~\ref{fig:azimuthal_scatter_summary}). The median transition redshift, after which the radial variation dominates, is $z \approx 0.9$ ($t_{\rm lookback} \approx 7.4 \Gyr$) at $R = 4 \kpc$, $z \approx 0.8$ ($t_{\rm lookback} \approx 6.9 \Gyr$) at $R = 8 \kpc$, and $z \approx 0.6$ ($t_{\rm lookback} \approx 5.8 \Gyr$) at $R = 12 \kpc$.

While we in general find no systematic differences between our \ac{LG}-like hosts and our isolated hosts, the transition redshift of the LG-like simulations is systematically higher than that of the isolated simulations at large radii.  At $R = 12 \kpc$ the the median transition of the \ac{LG} suite occurs at $z \approx 0.8$ ($t_{\rm lookback} \approx 6.9 \Gyr$) compared to $z \approx 0.5$ ($t_{\rm lookback} \approx 5.1 \Gyr$) for the isolated hosts. The overall distribution of transition redshifts for the isolated hosts falls within the distribution of transition redshifts for the \ac{LG} suite, with the $68$th percentiles almost entirely overlapping. The slight tendency for LG-like hosts to transition to radial domination earlier is unsurprising, given the results in \citet{Santistevan20}, who found that the main progenitors of our \ac{LG}-like galaxies formed earlier than those of the isolated galaxies. This earlier formation may be responsible for the earlier transition redshift as the disks undergo more orbital times, which smooths azimuthal variation, and have a longer time to build up strong radial gradients.

\begin{figure}
    \centering
    \includegraphics[width = \columnwidth]{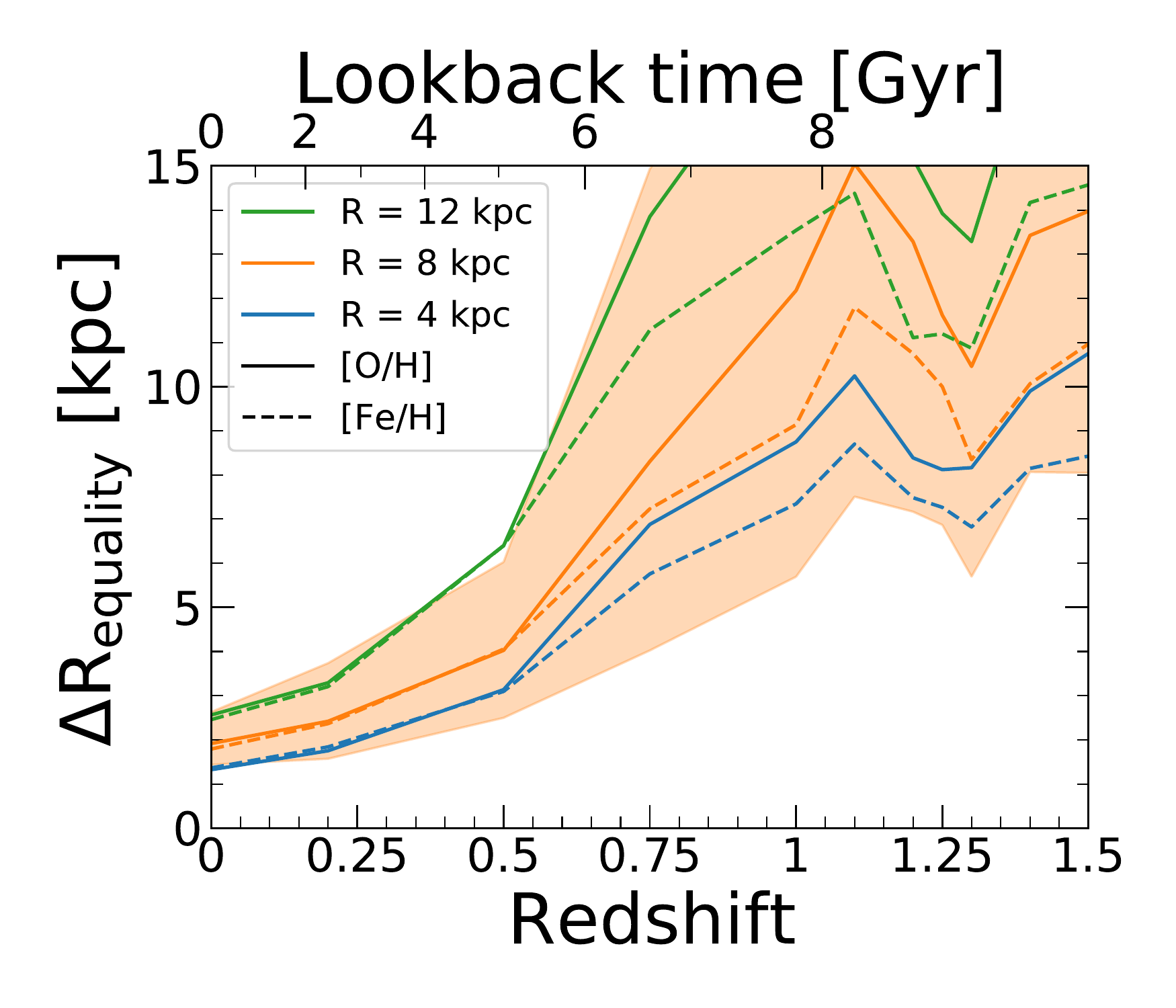}
    \vspace{-8 mm}
    \caption{
    $\Delta \rm R_{\rm equality}$ is the ratio of the azimuthal scatter to the radial gradient: it indicates the characteristic radial scale below which azimuthal variations dominate over radial variations, for \OH{} (solid) and \FeH{} (dashed).
    We compute the azimuthal scatter using the full ($360^\circ$) annulus of gas at $R = 4$, $8$, and $12 \kpc$. Lines show the median, and shaded regions show the $68$th percentile across our 11 galaxies.
    At early times, $\Delta \rm R_{\rm equality}$ is comparable to the size of the gas disk, when azimuthal scatter dominated over radial variation, but after $z \approx 0.5$ ($t_{\rm lookback} = 5.1 \Gyr$) the disk is well mixed azimuthally. At $z = 0$ the median $\Delta \rm R_{\rm equality}$ near the solar circle is $\approx 1.8 \kpc$ for \OH{} and $\approx 1.7 \kpc$ for \FeH, while at $z = 1.5$ ($t_{\rm lookback} = 9.4 \Gyr$) it is $\approx 14 \kpc$ for \OH{} and $\approx 11 \kpc$ for \FeH.
    For radial scales less than $\Delta \rm R_{\rm equality}$, the primary source of inhomogeneity of elemental abundances in gas is azimuthal variations, which complicate/limit the use of elemental abundances to infer a star's birth location to a radial precision smaller than $\Delta \rm R_{\rm equality}$.
    }
    \label{fig:azimuthal_radial_ratio}
\end{figure}

Fig.~\ref{fig:azimuthal_radial_ratio} shows another metric for comparing the importance of azimuthal versus radial variations across time.
For each simulation, we compute $\Delta \rm R_{\rm equality}$, the ratio of the $360^\circ$ azimuthal scatter (at a given radius) to the radial gradient.
This represents the characteristic radial scale over which the radial and azimuthal abundance variations are equal.
The lines show the median $\Delta \rm R_{\rm equality}$ for \OH{} and \FeH, measuring the azimuthal variation at 3 radii, and the shaded region shows the $68$th percentile of \OH{} at $R = 8 \kpc$. We apply boxcar smoothing to the data to minimize the significant stochasticity in these values at early times, when the radial gradient fluctuated over short timescales (see Sec.~\ref{subsec:radial_gradient_evolution}).

The median $\Delta \rm R_{\rm equality}$ for \OH{} is $\lesssim 14 \kpc$ at $z = 1.5$ ($t_{\rm lookback} = 9.4 \Gyr$) and $\lesssim 1.8 \kpc$ at $z = 0$ at the solar circle.  This corresponds to the radial range over which azimuthal scatter dominates the variations in abundance, rather than the radial gradient. Thus, for the purposes of chemical tagging, this represents a limit for the radial precision that chemical tagging (of a single element) can place on the formation location of a star without also modeling azimuthal location. At early times, $\Delta \rm R_{\rm equality}$ is comparable to or greater than the size of the disk, meaning that the azimuthal coordinate determines the abundance of newly forming stars more than the radial position.

$\Delta \rm R_{\rm equality}$ is largest at $z = 1.5$ ($t_{\rm lookback} = 9.4 \Gyr$) and then decreases with time, given the decreasing azimuthal scatter and increasing strength of the radial gradient with time. Also, over time the scatter across our 11 galaxies decreases. The high scatter at high redshifts is a result of the large scatter in both radial gradients and azimuthal variations at these times, given scatter in formation history.

Fig.~\ref{fig:azimuthal_radial_ratio} also shows the dependence of $\Delta \rm R_{\rm equality}$ on radius. This ratio slightly increases as a function of radius, because the azimuthal scatter increases with radius at all times (Fig.~\ref{fig:azimuthal_scatter_summary}).
This means that modeling chemical tagging of stellar birth radius is more challenging at larger radii.

In summary, any models for chemical tagging should incorporate azimuthal scatter in abundance especially at $z \gtrsim 0.6$ ($t_{\rm lookback} \gtrsim 5.8$), because the azimuthal scatter in gas dominates at these early times.

\subsection{Radial scale of measurable homogeneity}

We next explore observational implications of our measured radial gradients, by comparing them against typical measurement uncertainties in elemental abundances, to understand the radial scales over which gas is effectively homogeneous in a measurable sense.
Thus, in this sub-section we ignore azimuthal variations and focus just on radial gradients.
While we examine gas-phase abundances, the chemical tagging of stars (that form out of this gas) ultimately motivates our work, so we examine measurement uncertainties typical of \ac{MW} stellar surveys.

\begin{figure}
    \centering
    \includegraphics[width=\columnwidth]{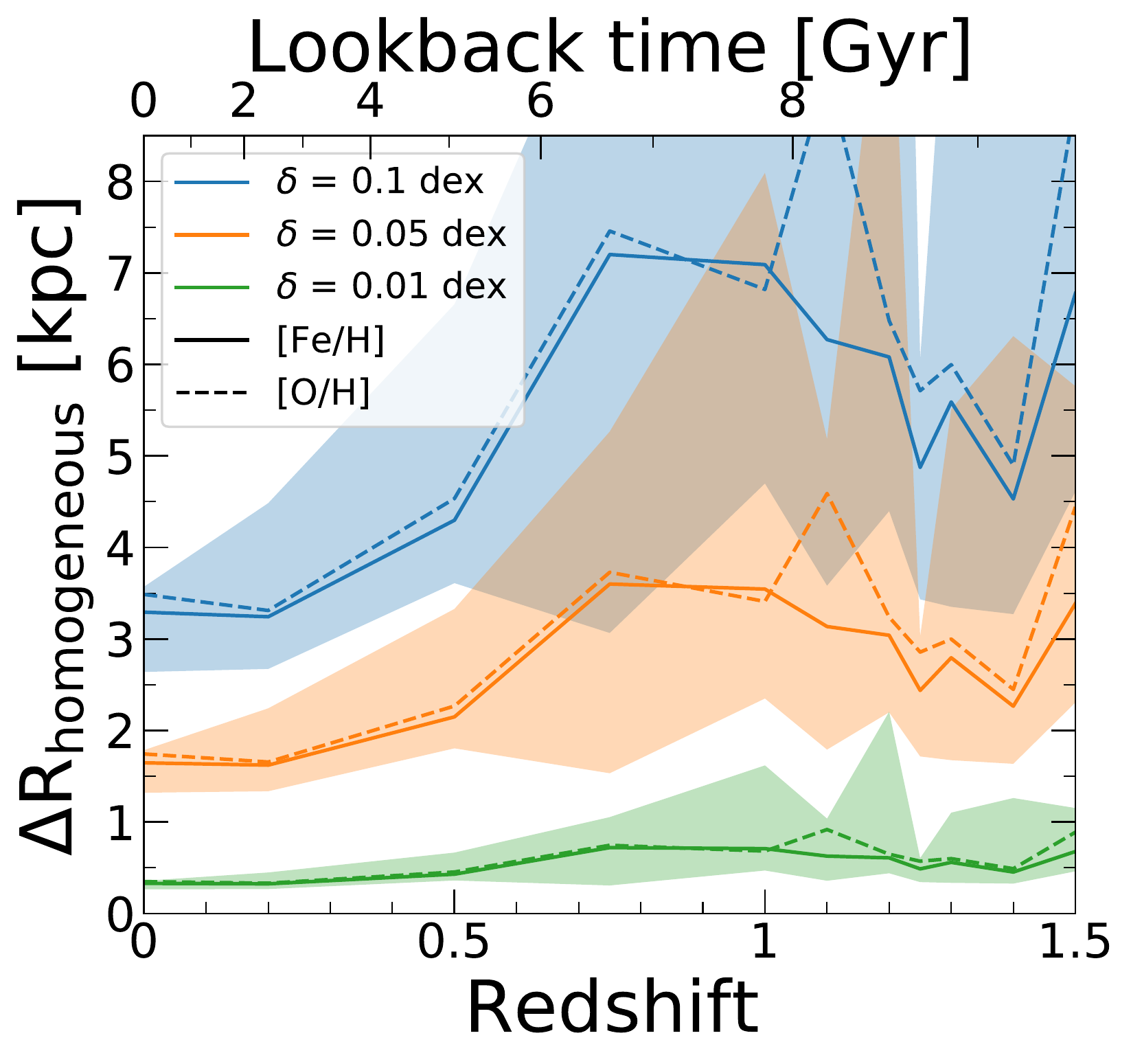}
    \vspace{-7 mm}
    \caption{
    Assuming a given measurement uncertainty in an elemental abundance, $\Delta \rm R_{\rm homogeneous}$ is the ratio of this measurement uncertainty to the radial gradient in that abundance; it is the characteristic radial scale below which the gas disk is effectively radially homogeneous in a measurable sense.
    We show $\Delta \rm R_{\rm homogeneous}$ versus redshift, for $3$ measurement uncertainties, motivated by observational surveys of stellar abundances.
    The lines show the median across our 11 galaxies for \FeH{} (solid) and \OH{} (dashed), and the shaded regions show the $68$th percentiles for \FeH.
    At $z = 1.5$ ($t_{\rm lookback} = 9.4 \Gyr$) the gas disk is measurably homogeneous (for $\delta = 0.05 \dex$) across significant radial scales, $\Delta R \lesssim 3 \kpc$, though this decreases to $\lesssim 1.6 \kpc$ at $z = 0$. For high-precision abundance measurements ($\delta = 0.01 \dex$), the gas disk is measurably homogeneous at
    $\Delta R \lesssim 0.6 \kpc$ at $z = 1.5$ and
    $\Delta R \lesssim 0.3 \kpc$ at $z = 0$.
    This highlights the limitations from just measurement uncertainties on chemical tagging to infer a star's birth radius; it does not include the additional complications from azimuthal variations (Fig.~\ref{fig:azimuthal_radial_ratio}), which can be more important.
    }
    \label{fig:obs_radial_ratio}
\end{figure}

Motivated by observational surveys of stellar abundances, we select 3 observational measurement uncertainties of $\delta_{\rm m} = 0.01$, $0.05$, and $0.1 \dex$.
Fig.~\ref{fig:obs_radial_ratio} shows $\Delta \rm R_{\rm homogeneous}$, the ratio of $\delta_{\rm m}$ to the radial gradient in abundance, versus redshift.
Unlike Fig.~\ref{fig:azimuthal_radial_ratio}, which shows radial scales of homogeneity at different radii, which depends on the azimuthal scatter at each radius, Fig.~\ref{fig:obs_radial_ratio} depends only on the radial gradient measured across the whole disk. Thus the radial scale in Fig.~\ref{fig:obs_radial_ratio} represents the average radial scale of observed homogeneity based on an azimuthally averaged radial gradient.
The solid line shows the median and the shaded region shows the $68$th percentile across our 11 hosts.
$\Delta \rm R_{\rm homogeneous}$ represents the radial scale of measurable homogeneity: below this radial length scale, the change in abundance from the radial gradient is less than this measurement uncertainty.
The larger $\Delta \rm R_{\rm homogeneous}$ is, the less precisely measurements can pinpoint a star's birth radius.

For a fiducial measurement uncertainty of $\delta = 0.05 \dex$ at $z = 1.5$ ($t_{\rm lookback} = 9.4 \Gyr$) $\Delta \rm R_{\rm homogeneous} \approx 3.1 \kpc$, which drops to $\approx 1.6 \kpc$ at $z = 0$ for \FeH{}. \OH{} is consistent to within $\approx 1 \kpc$ at all redshifts, except for $z = 1.5$. At early times, $\Delta \rm R_{\rm homogeneous}$ is larger and has large scatter.  The largest scatter, at $z = 1.5$, comes from the radial gradients being flattest: some galaxies have gradients approaching $0 \dpk$, as Fig.~\ref{fig:radial_grad_evolution} shows.
Fig.~\ref{fig:obs_radial_ratio} shows essentially an inverse gradient, so the short time fluctuations at early times (see Sec.~\ref{fig:radial_grad_evolution}) also lead to rapid and significant variations in $\Delta \rm R_{\rm homogeneous}$.
After $z = 0.75$ ($t_{\rm lookback} = 6.6 \Gyr$) $\Delta \rm R_{\rm homogeneous}$ decreases over time.  This means that chemical tagging with measured abundances can identify the birth radius of more recently formed stars more precisely.

Comparing Fig.~\ref{fig:azimuthal_radial_ratio} with Fig.~\ref{fig:obs_radial_ratio} shows that, in terms of limitations on chemical tagging for a star's birth radius, at $z \gtrsim 0.5$ ($t_{\rm lookback} \gtrsim 5.1 \Gyr$) azimuthal variations dominate over observational uncertainties in the inner disk, for a fiducial uncertainty of $\delta = 0.05 \dex$.  In the outer disk ($R \geq 8 \kpc$) azimuthal variations are larger than observational uncertainties for all redshifts.  For higher-precision measurements, $\delta = 0.01 \dex$, azimuthal variations dominate at all times at all radii. This implies that, if a primary motivation for chemical tagging is inferring the birth location of a star, there is not much benefit in pushing to higher precision, because azimuthal variations dominate. In fact, Fig.~\ref{fig:radial_v_azimuthal} can indicate the maximum precision in elemental abundance that one should aim to measure stars of a given age for this purpose, given our predicted azimuthal scatter, unless a given chemical tagging approach includes modeling azimuthal variations.
We will explore possible models in future work.

\subsection{Distributions of elemental abundances}

\begin{figure*}
    \includegraphics[width=\textwidth]{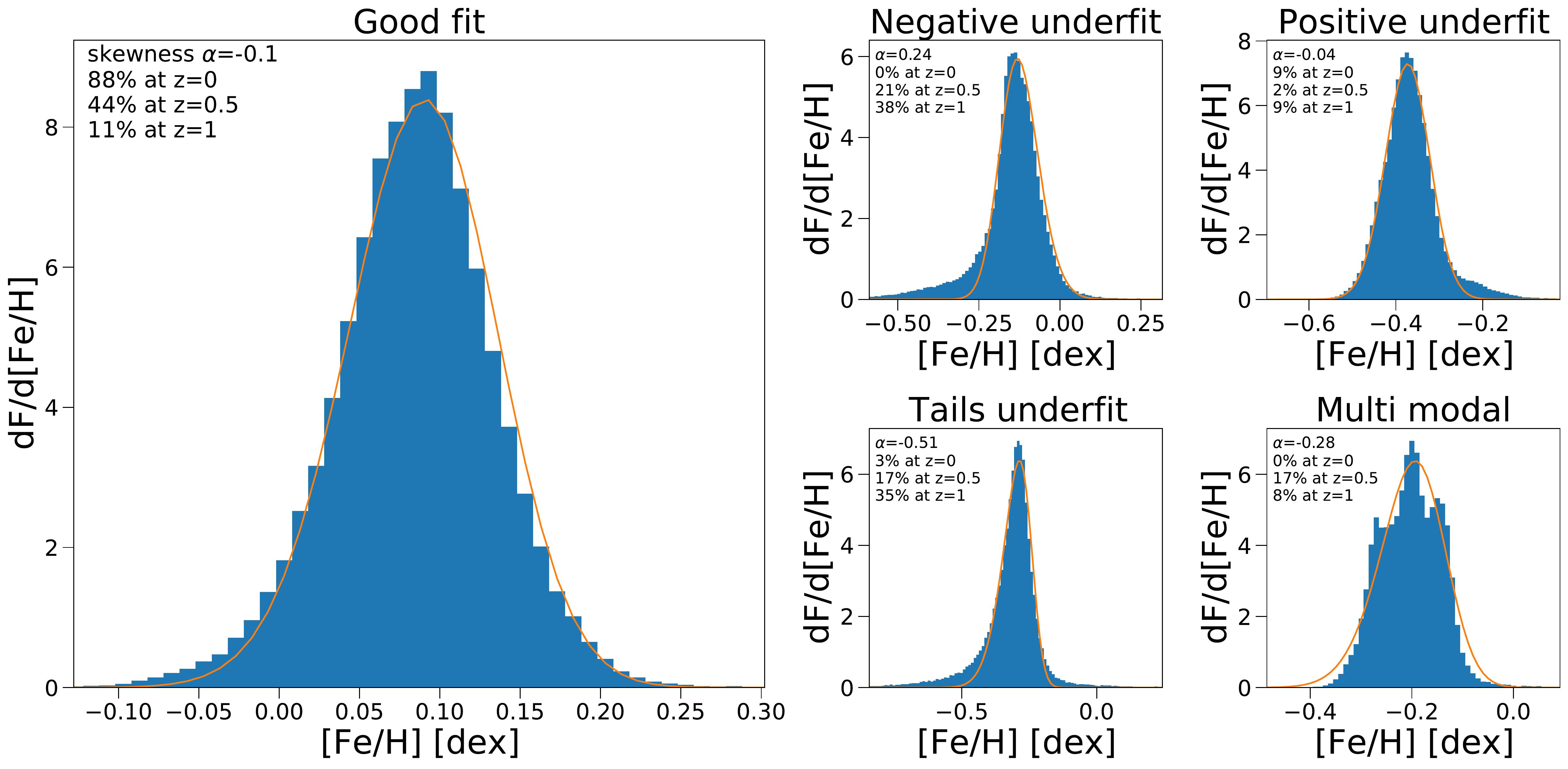}
    \vspace{-5 mm}
    \caption{
    Example distributions of \FeH{} from our 11 galaxies. Each panel shows the elemental distribution for a single simulation at a single radius. For each distribution, we stack 3 consecutive snapshots ($\Delta t \approx 50 \Myr$) and measure all gas within $\pm 0.2 \kpc$ of 3 radii ($R = 4$, $8$, and $12 \kpc$) and within a height $\pm 1 \kpc$ of the disk.  The solid line shows a skew normal fit to the distribution.  The left panel shows a well fit skew normal distribution. The right panel shows typical failure modes of fitting a skew normal distribution: underfitting negative skewness, underfitting positive skewness, underfitting the tails of the distribution, and multi modal distributions.  Each panel shows the fitted skewness for the example distribution along with the percentage of fits that fall into the category at each redshift across all radii and for both \OH{} and \FeH. In general, the simulations are well fit by a skew normal at $z = 0$, but it provides a worse fit at higher redshift, when the (negative) tails are preferentially underfit.  In general, the fit to the distribution of \OH{} and \FeH{} for the same $R$ and redshift fall into the same category.
    }
\label{fig:example_mdfs}
\end{figure*}

Finally, we explore the full distributions of elemental abundances in gas that our simulations predict.
Again, we emphasize that our \ac{FIRE}-2 simulations explicitly model the sub-grid diffusion/mixing of elemental abundances in gas via unresolved turbulent eddies, which is necessary to match observed distributions of abundances in the local Universe \citep{Su17, Escala18, Hopkins18}.

We measure \OH{} and \FeH{} distribution at $R = 4 \kpc$, $R = 8 \kpc$, and $R = 12 \kpc$ at $z = 1$ ($t_{\rm lookback} = 7.8 \Gyr$), $z = 0.5$ ($t_{\rm lookback} = 5.1 \Gyr$), and $z = 0$ for all galaxies. We fit these with a skew normal distribution, using the LevMarLSQ fitter in Astropy \citep{astropy:2013, astropy:2018}:
\begin{equation}
    \frac{dF}{dx} = A \times \exp{\left( -0.5 \left( \frac{x - \mu}{\sigma} \right)^2 \right)} \times \frac{1 + erf\left(\alpha \times \frac{x - \mu}{\sqrt{2}\sigma} \right)}{2}
    \label{eq:skew_normal}
\end{equation}
where $\mu$ is the mean, $\sigma$ is the standard deviation, and $\alpha$ is the skewness.
Fig.~\ref{fig:example_mdfs} shows representative example distributions of \FeH{}, for good and bad fits to this distribution, for a single simulation, and we list the percent of galaxies and radii that fall into each category.

As the left panel of Fig.~\ref{fig:example_mdfs} shows, a skew normal distribution reasonably fits these distributions in most cases at $z \sim 0$.
However, there are several common failure modes.  We categorize them as: failing to capture the positive or negative tails of the distribution, failing to capture the width of the distribution, or the distribution being multi-modal.
The right panel of Fig.~\ref{fig:example_mdfs} shows examples of each of these failures, along with the percentage of fits (\FeH{} and \OH{} combined) that we identify to fall into each category at each redshift, stacking all galaxies and radii at that redshift.
In general, the fit to \OH{} and \FeH{} at a given redshift and radius falls into the same category.  At $z = 0$, the vast majority ($\approx 88\%$) of the distributions are well fit.
The most common failure is a positive underfit, given pockets of high metal enhancement from feedback.
However, at $z = 1$ ($t_{\rm lookback} = 7.8 \Gyr$) the failures are more common, and only $\approx 11 \%$ are well fit.
Most common is having a negative underfit or both tails underfit, likely driven by more rapid accretion of low-metallicity gas at earlier times.

\begin{figure*}
	\includegraphics[width=\textwidth]{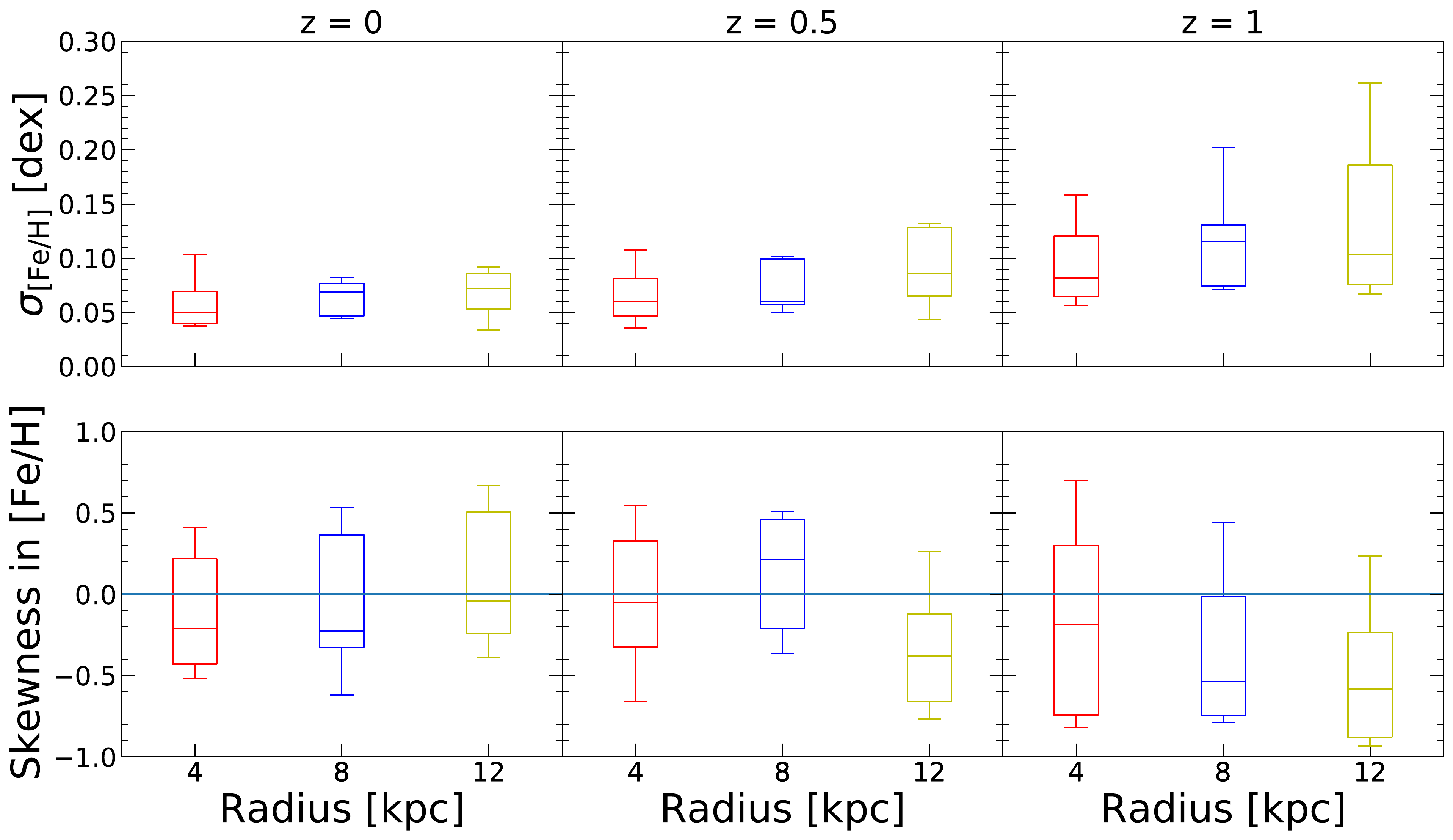}
	\vspace{-5 mm}
    \caption{
    \textbf{Top}: standard deviation of the fitted elemental distribution of gas \FeH{} at 3 radii, for increasing redshift (left to right). For each distribution, we stack 3 consecutive snapshots ($\Delta t \approx 50 \Myr$) and measure all gas within $\pm 0.2 \kpc$ of each radius and within a height $\pm 1 \kpc$ of the disk. Each box shows the $68$th percentile of the distribution and the whiskers show the full distribution of standard deviations. The standard deviation increases with radius  and decreases with time.
    \textbf{Bottom}: skewness of the fitted elemental distributions of gas \FeH{}.
    At earlier times, the gas disk had stronger negative skewness, but the disk relaxes to near zero skewness at $z = 0$. The skewness shows a slight radial dependence at both $z = 1$ ($t_{\rm lookback} = 7.8 \Gyr$) and $z = 0$.  At $z = 1$ the distributions at larger radii were more negatively skewed, whereas at $z = 0$ the distributions at smaller radii are more negatively skewed.
    }
    \label{fig:MDF_summary}
\end{figure*}

While not perfect, especially at earlier times, a skew normal fit does provide a simple characterization of the full distribution.
Thus, Fig.~\ref{fig:MDF_summary} shows the fit parameters for \FeH{} (\OH{}, not shown, is consistent with this) at different radii and redshifts. The box-and-whisker plots show the median, $68$th percentile, and the full distribution. The top row shows the fitted standard deviation, while the bottom row shows the fitted skewness, and left to right shows increasing redshift.

At all radii, $\sigma$ decreases over time. At $R = 8\kpc$, near the solar circle, the median standard deviation decreases from $\approx 0.12 \dex$ at $z = 1$ ($t_{\rm lookback} = 7.8 \Gyr$) to $\approx 0.07$ at $z = 0$ for \FeH{} ($\approx 0.11$ to $\approx 0.06$ for \OH).  Thus, consistent with the results for azimuthal scatter, gas at a given radius becomes more homogeneous over time.
Also consistent with our results for azimuthal scatter, Fig.~\ref{fig:MDF_summary} shows that at all redshifts, $\sigma$ increases with radius, that is, metals are less well mixed at larger radii. At $z = 0$ the median is $\sigma \approx 0.05$ at $R = 4 \kpc$ and increases to $\approx 0.07$ at $R = 12 \kpc$ for \FeH{} ($\approx 0.04$ to $\approx 0.06$ for \OH).

Fig.~\ref{fig:MDF_summary} (bottom row) shows that the distributions are preferentially negatively skewed at earlier times, but they trend toward Gaussian over time. At $R = 8 \kpc$ the median skewness is $\alpha \approx -0.54$ at $z = 1$ ($t_{\rm lookback} = 7.8 \Gyr$) and $\alpha \approx -0.21$ at $z = 0$ for \FeH{} ($\alpha \approx -0.52$ to $\approx -0.23$ for \OH). 
At earlier times, skewness decreases with radius, from $\approx -0.19$ at $R = 4\kpc$ to $\approx -0.58$ at $z = 1$ ($\approx -0.15$ to $\approx -0.5$ for \OH).
At earlier times, higher rates of cosmic accretion of pristine gas can skew the distributions negatively, especially at large radii, where enrichment also is more stochastic given lower star-formation rates and orbital/mixing times are longer. At later times, as the gas accretion and star-formation rates decrease, the distributions tend toward Gaussian, as metals become better mixed within each annulus.
At $z \sim 0$, all radii show abundance distributions consistent with no skewness at the $1 \mhyphen \sigma$ level.

\section{Summary and Discussion}
\label{sec:sum}

\subsection{Summary}

We use a suite of FIRE-2 cosmological zoom-in simulations of 11 \ac{MW}/M31-mass galaxies to explore the 3-D spatial variations and evolution of elemental abundances \OH{}, \FeH{}, and \NH{} in gas at $z \leq 1.5$ ($t_{\rm lookback} \leq 9.4 \Gyr$), to understand the birth conditions of stars to inform the efficacy of chemical tagging of stars.  While many stars form prior to $z = 1.5$, the last $\sim 10 \Gyr$ mark the primary epoch of disk assembly, which is where we are primarily interested in chemically tagging stars. Our main results are:

\begin{itemize}
    \item \textit{Vertical gradients}: are negligible. Abundances in gas are well mixed vertically at all times. At $R = 8 \kpc$, the mean deviation in \OH{} at $1 \kpc$ from the galactic midplane is $< 0.01 \dex$ at all times. The inner $\sim 200 \pc$ of the disks, where the majority of star formation for $z < 0.5$ occurs, is approximately uniform in abundance ($|\Delta \rm [O/H]| \lesssim 0.002 \dex$) at all times. The inner $\sim 1.5 \kpc$ of the disks, where the majority of star formation for $z > 0.5$ occurs has minimal vertical variation in abundance $|\Delta \rm [O/H]| \lesssim 0.01 \dex$ at all times. Thus there is minimal vertical information for chemical tagging.

    \item \textit{Radial gradients}: are negative at all times and for all abundances, with a maximum steepness of $\approx -0.03 \dpk$ at $z = 0$ and a minimum of $\approx -0.01 \dpk$ at $z \gtrsim 1$ ($t_{\rm lookback} \gtrsim 7.8 \Gyr$). Radial gradients become steeper over time, because the disks become more rotationally supported and are better able to sustain a gradient against radial mixing, as noted in analysis of FIRE-1 simulations in \citet{Ma17}. \NH{} has a steeper gradient at all times, because their production is dominated by stellar winds, whose mass-loss rates increase with metallicity in our simulations, enhancing the discrepancies between metal-rich and metal-poor regions.  \OFe{} shows little variation with redshift, and is approximately flat across the disk indicating it provides limited discriminating power for chemically tagging birth radii.
    Our \OH{} gradients broadly agree with most observations of nearby \ac{MW}-mass galaxies, including M31, at the $1$ or $2 \mhyphen \sigma$ level, though our gradients are somewhat steeper on average. By contrast, our gradients are somewhat shallower than most observations of the MW, though they agree at the $1$ or $2 \mhyphen \sigma$ level with 9 of 13 \ac{MW} observations. 

    \item \textit{Azimuthal scatter}: systematically decreases over time for all abundances, from $\approx 0.2 \dex$ at $z = 1.5$ ($t_{\rm lookback} = 9.4 \Gyr$) to $\approx 0.05 \dex$ at $z = 0$ for \OH{} and \FeH{} around the entire disk at $R = 8 \kpc$.
    This evolution is a result of higher gas accretion and also star-formation rates at earlier times, which lead to stronger variations in abundances on small scales, especially at larger radii, where orbital/mixing timescales are longer. The azimuthal scatter in \NH{} is larger (by $\approx 0.01 \dex$ at $R = 8 \kpc$) at all times than in \FeH{} or \OH{}, for the same reasons as above.
    Azimuthal variations reduce somewhat with smaller azimuthal aperture. However, even in angular bins as small as $\approx 350 \pc$, they remain $\approx 0.04 \dex$ at $z = 0$ and $\approx 0.1 \dex$ at $z = 1$ ($t_{\rm lookback} = 7.8 \Gyr$) for \OH{} and \FeH. We emphasize that our azimuthal bins do not center on \ac{GMC}s or star-forming regions, so our results probe the homogeneity of effectively random patches of gas.
    We find good agreement between our azimuthal scatter in \OH{} in gas at $z = 0$ ($\approx 0.05 \dex$) and observations of nearby galaxies \citep{Sakhibov18, Kreckel19, Kreckel20}.

    \item \textit{Azimuthal versus radial scatter}: At early times, the azimuthal scatter was larger than the radial variation for all abundances. We quantify the redshifts when the radial variation (across $\Delta R = 8 \kpc$) first dominates over the azimuthal scatter, finding a median of $z \approx 0.9$ ($t_{\rm lookback} \approx 7.4 \Gyr$) at $R = 4 \kpc$ and $z \approx 0.6$ ($t_{\rm lookback} \approx 5.8 \Gyr$) at $R = 12 \kpc$.
    Before this time, stars born at the same radius could have the same difference in metallicity as stars born $\Delta R \gtrsim 8 \kpc$ apart.
    We also quantify across time the radial range over which the radial and azimuthal variations are comparable, $\Delta R_{\rm equality}$.
    At $z \sim 0$, $\Delta R_{\rm equality}$ is small at $\approx 1.8 \kpc$, but at $z \gtrsim 1$ ($t_{\rm lookback} \gtrsim 7.8 \Gyr$) $\Delta R_{\rm equality}$ is larger than the size of the disk.
    These results indicate that azimuthal variations in abundances provide the \textit{dominant} information content for chemical tagging for stars formed $\gtrsim 6 \Gyr$ ago, so future approaches to chemical tagging of stars should start to incorporate/model these significant azimuthal variations.

    \item \textit{Measurable homogeneity}: 
    We quantified the radial scales across which our gas disks are effectively homogeneous in a measurable sense, given representative measurement uncertainties.
    For an uncertainty in elemental abundance of $0.05 \dex$, our gas disks are measurably homogeneous across $\Delta R \approx 1.7 \kpc$ at $z = 0$ and $\Delta R \approx 3.5 \kpc$ at $z \gtrsim 0.75$ ($t_{\rm lookback} \gtrsim 6.6 \Gyr$).
    Moreover, azimuthal variations at $R \gtrsim 8 \kpc$ are larger than $0.05 \dex$ at all times.
    Thus, for any measurement uncertainty at or below this, using chemical tagging to measure birth radius is limited not by measurement uncertainty but instead by azimuthal variations.
    These results inform the needed precision for observations, given targeted precision for chemical tagging of stars across age/time.
    For example, if one only cares about modeling birth radius, there is little-to-no benefit in measuring a stellar abundance to better than $\approx 0.05 \dex$.

    \item \textit{Elemental abundance distributions}: We measured the full distributions of elemental abundances in radial annuli and fit them with skew normal distributions.  The skew normal distributions fit these distributions reasonably well, but there are failure modes that become more common at higher redshift, most notably underfitting the negative tails of the distribution and simultaneously underfitting the positive and negative tails. We find typically negatively skewed normal distributions at $z \gtrsim 1$ ($t_{\rm lookback} \gtrsim 7.8 \Gyr$), with stronger negative skewness at larger radii. The distributions evolve toward approximately Gaussian distributions at all radii by $z = 0$.

\end{itemize}

\subsection{Discussion}

The primary goal of this paper is to understand the homogeneity of gas as a proxy for the birth conditions of stars across space and time, as a first step to understanding the efficacy of chemical tagging in a cosmological context. There are caveats to our analysis, though. Namely, our analysis is performed looking at individual elements, with the exception of \OFe. Examining multi-element abundance distributions may well offer more discerning power. Our analysis of \OFe{} suggests that this may be limited. Furthermore, uncertainty in our fiducial diffusion coefficient leads to uncertainty in our small scale azimuthal abundance scatter, as seen in Appendix.~\ref{sec:diffusion_coefficient_test}.

Additional complications arise when comparing our simulations to observational data. We present results in the context of constraining chemical tagging in the MW, but our simulations are not exact MW-analogs. Also, when comparing the redshift evolution of our results to observations of external galaxies, we track the evolutionary history of individual galaxies across time, as opposed to measuring properties of different galaxies at fixed mass across time. For all of our comparisons to observations, we explore all gas whereas observers typically measure abundances in HII regions specifically.  However, \cite{Hernandez21}  compared observations of ionized and neutral gas-phase abundance gradients in M$83$, finding gradients for neutral gas to be $-0.17 \dpk$ and gradients for ionized gas to be $-0.03 \dpk$.  This might imply that our measured gradients are much flatter than one would expect, given observations.  \citet{Hernandez21} did the same analysis excluding the nuclear region of M$83$ and found the neutral gas to be in much better agreement with ionized gas (a gradient of $-0.02 \dpk$.  Additionally, we compare to observations which have measurements in broadly similar physical regions to those we analyze in the simulations, but they are not exactly the same.

We compared against observations of radial gradients in the \ac{MW}, M31, and similar-mass galaxies at $z = 0$, finding broad agreement.
We also connect our evolutionary trends with high-redshift observations of gas-phase abundances. In particular, we find that our \ac{MW}/M31-mass galaxies all have negative radial gradients at $z \sim 0$ but had nearly flat radial gradients at $z \gtrsim 1$ (where the average stellar mass of the hosts is $M_{90} \approx 1.74 \times 10^{10} \Msun$).
This trend agrees well with many observations of comparable mass galaxies at these higher redshifts \citep[e.g.][]{Queyrel12, Stott14, Wuyts16, Patricio19, Curti20}. However, some observational works have found strong negative radial gradients at these masses and redshifts \citep{Wuyts16, Carton18, Wang19}. Furthermore, while less common than negative radial gradients, some observations find some positive radial gradients at these redshifts as well \citep{Queyrel12, Wuyts16, Carton18, Wang19} which we do not find in any of our galaxies. In general, we find that the steepening of radial gradients with time in our simulations is consistent with observational results and follows the quantitative trends in other theoretical analyses, both in simulations \citep[e.g][]{Ma17} and in metallicity-evolution modeling \citep[][]{Sharda21}.

One of the most important aspects of our analysis is quantifying azimuthal variations in gas abundances and comparing their strength relative to radial gradients across cosmic time.
With the advent of integral field spectroscopy, observations have begun characterizing $2$-D abundance distributions in nearby galaxies.
These works \citep{Sanchez15, Vogt17, Ho17, Ho18} all find non-trivial azimuthal variations in nearby galaxies, for example, \citet{Kreckel19} found variations of $\approx 0.05 \dex$ at fixed radius, which agrees well with our results.
However, some observations \citep[e.g.][]{Zinchenko16} found no evidence for large-scale azimuthal variation in nearby galaxies.
One of our key results/predictions is the evolution of azimuthal variations, which we predict were stronger at higher redshifts.
Observations of gravitationally lensed systems now allow sub-kpc measurements at high redshift \citep{Jones13, Jones15}, making it possible to test this predicted evolution in more detail.

\citet{Kreckel20} examined azimuthal variations in gas-phase \OH{} across eight nearby galaxies using PHANGS-MUSE optical integral field spectroscopy.  While our technique for measuring azimuthal variations are not exactly comparable to their methods, we find similar results.  In our analysis we focus on scatter in all gas by measuring a mean scatter in angular bins of varying size, so we in effect measure the azimuthal inhomogeneities of random patches of gas at a given radius.  In contrast to this, \citet{Kreckel20} measure abundances specifically in HII regions and determine scatter by first subtracting off the radial gradient and then centering apertures of various sizes on individual HII regions and measuring the scatter between the HII regions contained within the aperture.  They find a slight scale dependence associated with the scatter, which we also see at $z = 0$, with the scatter on scales larger than $\approx 3 \kpc$ being $\approx 0.05 \dex$.  The small-scale scatter in \citet{Kreckel20} ($\approx 0.02 \dex$) is slightly smaller than the $z = 0$ scatter we observe, but this could be attributed to the discrepancy in our methods.  HII regions are likely better mixed in abundances than random patches of gas, so our analysis may be artificially inflating the typical azimuthal scatter of the gas from which stars are forming.  However, centering on HII regions is beyond the scope of our analysis, and in future work we will examine azimuthal variations in newly formed stars, which may be closer to the values in HII regions.

\citet{Krumholz18} derived the expected correlation function of metal distribution in galaxies across space and time using a stochastic diffusion model.  While we did not explore the correlation function of metals, we did examine homogeneity as a function of azimuthal scale, which we can compare broadly with their work. They found that gas-phase abundances produced primarily through core-collapse supernovae, in \ac{MW}-like conditions near the solar circle, are correlated on scales of $\approx 0.5 - 1 \kpc$ giving an expected scatter of $0.04 - 0.1 \dex$. From fully cosmological simulations our results for azimuthal scatter on scales of $\approx 1 \kpc$ near the solar cylinder agree well with their predicted range.

All of our results agree with \citet{Ma17}, who analyzed radial gradients of abundances in the FIRE-1 simulations across a much wider galaxy mass range.
In comparing with other theoretical/simulation works, our gradients in \OH{} at $z = 0$ fall between the gradients \citet{Hemler20} measured in the TNG50 simulations ($\approx -0.02 \dpk$) and the gradients \citet{Gibson13} measured in the MaGICC and MUGS simulations ($\approx -0.04 \dpk$).
In particular, \citet{Hemler20} found a gradual flattening of the gradients with time, which could come from an `inside-out' growth of galaxies wherein star formation, hence elemental enrichment, proceeds from the inner galaxy to the outer galaxy \citep[e.g.][]{Prantzos00, Bird13}. The flat(ter) radial gradients at earlier times in our galaxies result from higher turbulence and outflows that frequently eject much of the \ac{ISM} at those times, perturbations such as mergers and rapid gas infall result in the velocity dispersion of gas particles dominating over their rotational velocity leading to galaxy-scale radial mixing \citep{Ma17}. As the disk settles over time, it becomes more rotationally supported, so stronger radial gradients can develop/persist.
Our results qualitatively agree with those of the EAGLE simulations \citep{Tissera19}, though as with TNG50, \citet{Tissera19} found \OH{} gradients that are slightly shallower than ours, $\approx -0.011 \dpk$ at $z = 0$.

The evolution of our gas-phase abundance gradients disagrees with \citet{Agertz21}, who analyzed the VINTERGATAN simulation of the m12i initial conditions, performed using the \ac{AMR} code RAMSES. They found that the gas-phase profile of \FeH{} becomes shallower over time (their Fig.~7), compared with our steepening with time. One possible explanation is the difference in hydrodynamic solvers: we use the mesh-free finite-mass (MFM) quasi-Lagrangian method in \textsc{Gizmo}, coupled with explicit modeling of sub-grid mixing, while the AMR simulation of VINTERGATAN induces significantly more mixing in gas (complete mixing on the scale of an individual cell), which may contribute to the flattening of their gradient over time.  However, as shown in Appendix.~\ref{sec:diffusion_coefficient_test}, the qualitative steepening of the gradient we observe with time is independent of the strength of our diffusion coefficient.

Galactic evolution models often simplify the abundance distributions of gas in galaxies to a $1$-D model \citep[e.g.][]{Minchev18, Molla19a, Frankel20}, with azimuthal scatter assumed from measurements at $z = 0$. While this is a useful first step in understanding the abundance evolution of galaxies and testing chemical tagging, our results mean that this simplifying assumption overestimates the radial information content in elemental abundances, including how well chemical tagging can constrain the birth radius of a star. On the one hand, the non-trivial azimuthal scatter that we find, especially at earlier times, complicates modeling the abundances of stars at a given radius.
On the other hand, this likely makes individual \ac{GMC}s more elementally distinct at fixed radius, providing greater discriminating power, as we will explore in future work. However, we do not explore the homogeneity of individual \ac{GMC}s in this work.
Recent work has started to pursue $2$-D abundance evolution models \citep[e.g.][]{Molla19b} which may address this question, works such as \citet{Spitoni19} find azimuthal abundance variations on the order of $0.1 \dex$, twice what we find at $z = 0$ in our simulation suite.

Related to our analysis of a transition epoch, after which radial variations dominate over azimuthal scatter as the disk settles, \citet{Yu2021} examine the transition epoch from `bursty' to `steady' star formation and disk settling in the same simulations. We checked that 
their measurement of this bursty/steady transition agrees moderately well with the transition epoch that we present here, at least at smaller radius ($4 \kpc$). We find weaker agreement for our transition times at larger radii ($8$ and $12 \kpc$).
Furthermore, our transition times are consistently earlier ($\sim 3 \Gyr$ at $R = 4 \kpc$ and $\sim 1 \Gyr$ at $R = 12 \kpc$) than those in \citet{Yu2021}, with the transition times on average being more similar at larger radii.  Thus, we find a broad correlation between the transition from bursty to smooth star formation and the transition from azimuthal to radial abundance variations, but with significant scatter and some time delay.

Our simulations show the importance of considering azimuthal variations in addition to radial variations when studying gas-phase elemental abundance distributions.  This is particularly important in the context of chemical tagging; in order to accurately identify the birth locations of stars using elemental abundances the initial conditions of stars need to be well defined. As we showed in Section.~\ref{subsec:azimuthal_vs_radial} azimuthal variations in abundance are greater than or comparable to radial variations at earlier times, so chemical tagging models that only account for radial variations will fail to accurately capture the scatter in abundances at a given radius. This could lead to incorrectly assigning stars as co-natal, or vice versa.
We also fit the elemental distributions of our galaxies at different radii at different times, finding that they shift from negative to zero skewness over time. While skew-normal fits are not a perfect fit for the elemental distributions of our galaxies at all redshifts, they more accurately represent the distributions than a Gaussian. Thus, using our measured distributions would be a useful step in building more accurate abundance evolution models including for chemical tagging.

Next generation telescopes are crucial for testing the predictions of gas-phase abundance homogeneity presented in this work, particularly the predictions for azimuthal scatter at high redshifts. Current measurements of azimuthal scatter in abundances have been restricted to nearby galaxies. However, with the advent of JWST NIRSPec IFU and next-generation adaptive-optic IFUs on telescopes like IRIS and TMT, spatially resolved measurements of metallicities in distant galaxies are feasible, providing tests of our predictions for azimuthal scatter and the transition redshifts when it becomes sub-dominant.

This work is the first step of testing the limits of chemical tagging in the \ac{FIRE} simulations. In the future we will examine the degree to which these results for gas are mirrored in newly formed stars across time.
Combining those results with measurements of the dynamical evolution of stars in our simulations, we more directly will test the efficacy of chemical tagging of stars in the FIRE simulations.

\section*{Acknowledgements}
\label{sec:ack}

We thank Francesco Belfiore for providing abundance gradient measurements for galaxies from the MaNGA survey. We also thank Tucker Jones, Ryan Sanders, Joss Bland-Hawthorn, Trey Wenger, and Dana Balser for useful discussions that improved this manuscript.
We thank the anonymous reviewer for useful comments that improved this article.

MAB and AW received support from NASA through ATP grants 80NSSC18K1097 and 80NSSC20K0513; HST grants GO-14734, AR-15057, AR-15809, and GO-15902 from STScI; a Scialog Award from the Heising-Simons Foundation; and a Hellman Fellowship.
Support for SRL was provided by NASA through Hubble Fellowship grant \#HST-JF2-$51395.001$-A awarded by STScI, which is operated by AURA, Inc., for NASA, under contract NAS5-26555.
We ran simulations using XSEDE supported by NSF grant ACI-1548562, Blue Waters via allocation PRAC NSF.1713353 supported by the NSF, and NASA HEC Program through the NAS Division at Ames Research Center.  CAFG was supported by NSF through grants AST-1715216 and CAREER award AST-1652522; by NASA through grant 17-ATP17-0067; by STScI through grant HST-AR-16124.001-A; and by a Cottrell Scholar Award and a Scialog Award from the Research Corporation for Science Advancement.  RF acknowledges financial support from the Swiss National Science Foundation (grant no 194814).
We performed this work in part at the Aspen Center for Physics, supported by NSF grant PHY-1607611, and at KITP, supported by NSF grant PHY-1748958.

\section*{Data Availability}

Full simulation snapshots at $z = 0$ are available for m12i, m12f, and m12m at ananke.hub.yt.
A public version of the GIZMO code is available at http://www.tapir.caltech.edu/~phopkins/Site/GIZMO.html. Additional data including simulation snapshots, initial conditions, and derived data products are available at https://fire.northwestern.edu/data/.
Some of the python code used to analyze these data includes the publicly available packages https://bitbucket.org/awetzel/gizmo\_analysis \citep{GIZMO20} and https://bitbucket.org/awetzel/utilities.




\bibliographystyle{mnras}
\bibliography{PaperDraft} 




\appendix

\section{Scaled radial profiles} 
\label{sec:scale_lengths}

\begin{figure}
    \centering
    \includegraphics[width=\linewidth]{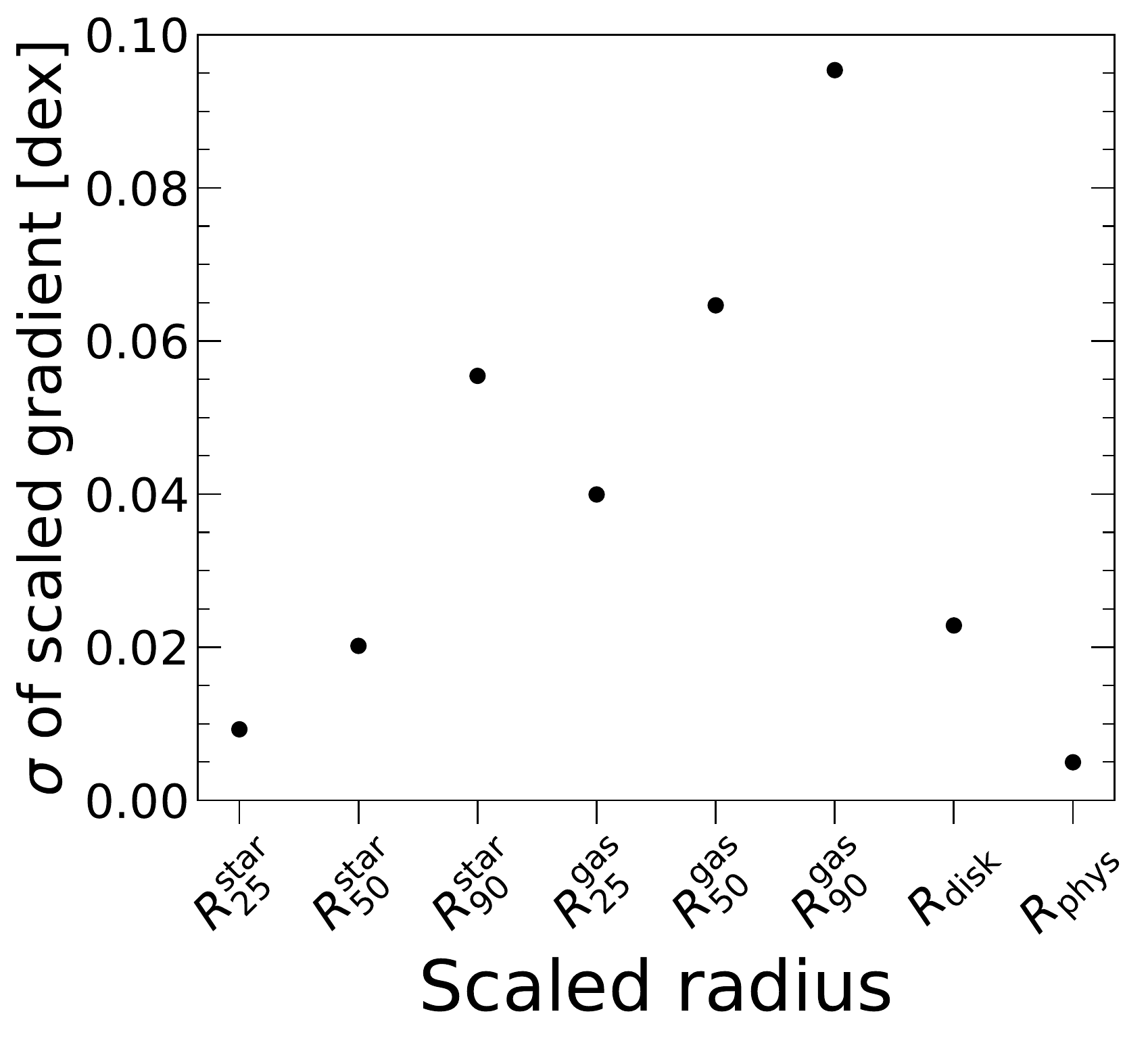}
    \vspace{-6 mm}
    \caption{
    The $1 \mhyphen \sigma$ host-to-host scatter of the scaled radial gradients for different scale radii at $z = 0$. We define $R_{25}$, $R_{50}$, and $R_{90}$ for the gas and stars in Table~\ref{table:scale_lengths}.
    $R_{\rm disk}$ is the exponential scale length of the stellar disk determined via a 2-component fit to the surface density, and $R_{\rm phys}$ is the physical radial coordinates of the disk, i.e. unscaled coordinates.  For each scale radius, we measure the gradient of all galaxies across an equal radial range that corresponds to $4 - 12 \kpc$ physical for the galaxy with the median scale length. The $1 \mhyphen \sigma$ host-to-host scatter is smallest when measuring the gradients in physical space, which motivates our choice for our analysis in this paper.
    }
    \label{fig:scaled_gradient_scatter}
\end{figure}

Fig.~\ref{fig:scaled_gradient_scatter} compares the host-to-host scatter in radial gradients of  \OH{} in gas in our simulated galaxies when scaling these gradients to various galaxy scale radii at $z = 0$.
We scale each galaxy's profile using: $R_{25}$, $R_{50}$, and $R_{90}$ for the gas and the stars, along with the exponential scale length, $R_{\rm disk}$, from a 2-component (s\'{e}rsic plus exponential) fit to the surface density.
Table~\ref{table:scale_lengths} lists the values for each galaxy.
We also compare these scaled gradients to the gradients in physical radii (as measured in Section.~\ref{subsec:radial_gradient_evolution}.
We bin each profile equally in scaled radius, defining the bin width such that the galaxy with the median scale length has a binwidth of $0.25 \kpc$ physical.
We measure the radial gradient of each galaxy across an equal radial range for each scale radius. We define this radial range such that we measure the galaxy with the median scale length across a physical range $4 - 12 \kpc$.
This range corresponds to: $\approx 3.3 - 9.8 R^{\rm star}_{25}$, $\approx 1.4 - 4.1 R^{\rm star}_{50}$, $\approx 0.4 - 1.1 R^{\rm star}_{90}$, $\approx 0.5 - 1.6 R^{\rm gas}_{25}$, $\approx 0.3 - 1.0 R^{\rm gas}_{50}$, $\approx 0.2 - 0.7 R^{\rm gas}_{90}$, and $\approx 1.0 - 3.1 R_{\rm disk}$.

Measuring the gas abundance radial gradient (from $4 - 12 \kpc$) in physical space minimizes the host-to-host scatter, to $\sigma \approx 0.005 \dex$. $R^{\rm star}_{25}$ has the next smallest $1 \mhyphen \sigma$ scatter with $\sigma \approx 0.009 \dex$. The gradients are the least self-similar when scaled by $R^{\rm gas}_{90}$, for which the $1 \mhyphen \sigma$ scatter is $\approx 0.1$.
The self-similarity of the radial profiles in physical space motivates our choice in this paper, because there is no compelling reason to scale the profiles of our galaxies. We emphasize, though, that this may be a result of the small mass range of our suite (halo masses are $M_{\rm 200m} = 1 - 3 \times 10^{12} \Msun$, stellar masses are in Table~\ref{table:scale_lengths}) and may not be generalizable to galaxies across a wider mass range.

\section{All gas versus star-forming gas}
\label{subsec:all_v_star_forming_gas}

\begin{figure*}
    \includegraphics[width = \linewidth]{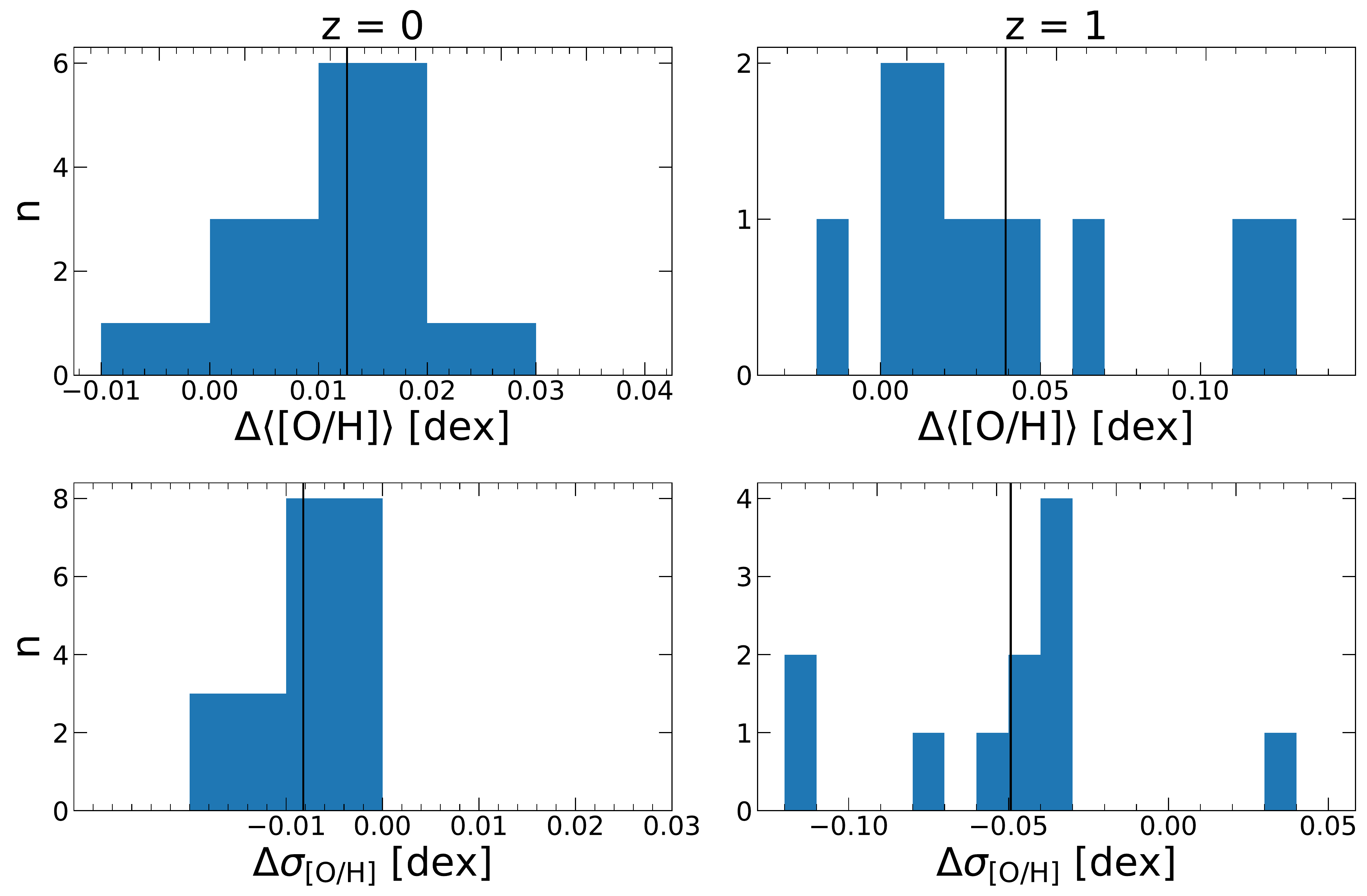}
    \vspace{-5 mm}
    \caption{
    A comparison of \OH{} measured in all gas versus only star-forming gas, at $4 < R < 12 \kpc$ and $|Z| < 1$ kpc. For each host, we measure the mean and standard deviation of its \OH{} (stacking 10 consecutive snapshot across $\approx 200 \Myr$ to boost the number of star-forming gas elements), and we compute the galaxy-wide difference between star-forming gas and all gas.
    Top panels show histograms of the difference in the mean \OH{}, while bottom panels show histograms of the difference in the standard deviation of \OH{}. Left panels show $z = 0$ and right panels show $z = 1$ ($t_{\rm lookback} = 7.8 \Gyr$). The solid vertical lines show the mean of each difference. Star-forming gas is on average more metal rich than all gas by $\approx 0.04 \dex$ at $z = 1$ and $\approx 0.01 \dex$ at $z = 0$. Furthermore, star-forming gas is slightly better mixed (with less scatter), with $\sigma_{\rm [O/H]} \approx 0.05 \dex$ smaller at $z = 1$ and $\approx 0.008 \dex$ smaller at $z = 0$.
    }
    \label{fig:OonH_all_vs_sf_summary}
\end{figure*}

In this paper, we examine elemental abundances in all gas, as initial conditions for chemical tagging of stars. We choose to measure all gas in part because star-forming gas represents only a small fraction of all gas elements at a given snapshot, leading to significant Poisson noise. In principal, we could attempt to identify photo-ionized (HII) regions near young star particles to compare with gas-phase measurements via nebular emission lines, but doing this correctly requires generating synthetic observations via ray-tracing, which is beyond the scope of our analysis.
In future work (Bellardini et al., in prep.) we will compare in detail the spatial variations in abundance of star particles that form out of this gas to the gas itself.
Here, we explore the impact of measuring only star-forming gas instead of all gas.

Fig.~\ref{fig:OonH_all_vs_sf_summary} compares measuring \OH{} in star-forming versus all gas at $z = 1$ ($t_{\rm lookback} = 7.8 \Gyr$) and $z = 0$. For each galaxy, we select gas elements at $4 < R < 12 \kpc$ and $|Z| < 1 \kpc$, and we stack this measurement across 10 snapshots ($\approx 200 \Myr$), because at any single snapshot there are few star-forming gas elements.
For reference, for these same simulations at $z = 0$, \citet{Benincasa20} find typical \ac{GMC} lifetimes, and hence lifetimes of star-forming regions, of $5 - 7 \Myr$.
We first measure the difference in the average abundance between star-forming and all gas for each galaxy.
Fig.~\ref{fig:OonH_all_vs_sf_summary} (top row) shows a histogram of this offset in the mean \OH. A positive value means star-forming gas has a higher \OH{} than all gas for that galaxy. The black vertical line shows the mean value of the histogram. On average, star-forming gas has modestly higher \OH{} than all gas by $\approx 0.04 \dex$ at $z = 1$ and $\approx 0.01 \dex$ at $z = 0$. The difference in \OH{} is typically $\lesssim 0.02 \dex$ for $z = 0$ and always less than $0.03 \dex$. The discrepancy is larger at higher redshift, the difference is typically $\lesssim 0.04 \dex$ and always less than $0.13 \dex$.
This is likely because cosmic accretion and star-formation rates are higher at earlier times, leading to less efficient small-scale mixing of gas.
Of course, a simple offset in the \OH{} normalization does not alone mean that spatial variations are different.

Fig.~\ref{fig:OonH_all_vs_sf_summary} (bottom row) shows the difference in the standard deviation of star-forming versus all gas.  Again, the black line shows the mean value.  On average, \OH{} for star-forming gas has slightly smaller standard deviation than for all gas. This difference is larger at $z = 1$ than at $z = 0$. However, the difference is typically small, $\lesssim 0.05 \dex$.
This suggests that the azimuthal variations of star-forming gas may be smaller than that of all gas, especially if the scatter is driven primarily by radial variations in abundance. Thus, chemically tagging the birth radii of stars may may be complicated by azimuthal variations for redshifts higher than we show in Sec.~\ref{subsec:azimuthal_vs_radial}, which we will explore further in Bellardini et al. in prep.

We also explore the differences in the radial gradients for star-forming versus all gas (not shown).
At $z = 0$ the radial gradients of star-forming and all gas are consistent to within $\pm 0.005 \dpk$, less than the host-to-host scatter in Fig.~\ref{fig:radial_grad_evolution}.
At $z = 1$, for the galaxies with sufficient star-forming gas to measure a reliable radial gradient, they agree with the gradients for all gas to within $\pm 0.002 \dpk$.
We also compare compare the radial profiles of \OH{} for newly formed stars (in age bins of $200 \Myr$) with that of all gas at the same snapshots. For $z \lesssim 0.5$, these profiles overlap to within uncertainty.
For $z \gtrsim 0.5$, the stellar and gas abundance profiles start to diverge, such that the young stars tend to have higher \OH{} and flatter gradients than all gas, which we will explore further in Bellardini et al., in prep.

In summary, while analyzing all gas is a reasonable, if not perfect, proxy for star-forming gas in our simulations. Given the short lifetimes of star-forming gas clouds \citep{Benincasa20} and the strict conditions for particles to be star-forming \citep{Hopkins18}, only a small fraction of gas elements are star-forming at a given snapshot ($\lesssim 1\%$ of gas particles at the redshifts we observe), so analyzing all gas greatly reduces the statistical uncertainty. In the future we will study the abundances of star particles that form from this gas, to compare with these results in detail.

\section{Impact of diffusion coefficient}
\label{sec:diffusion_coefficient_test}

\begin{figure*}
    \centering
    \includegraphics[width = \linewidth]{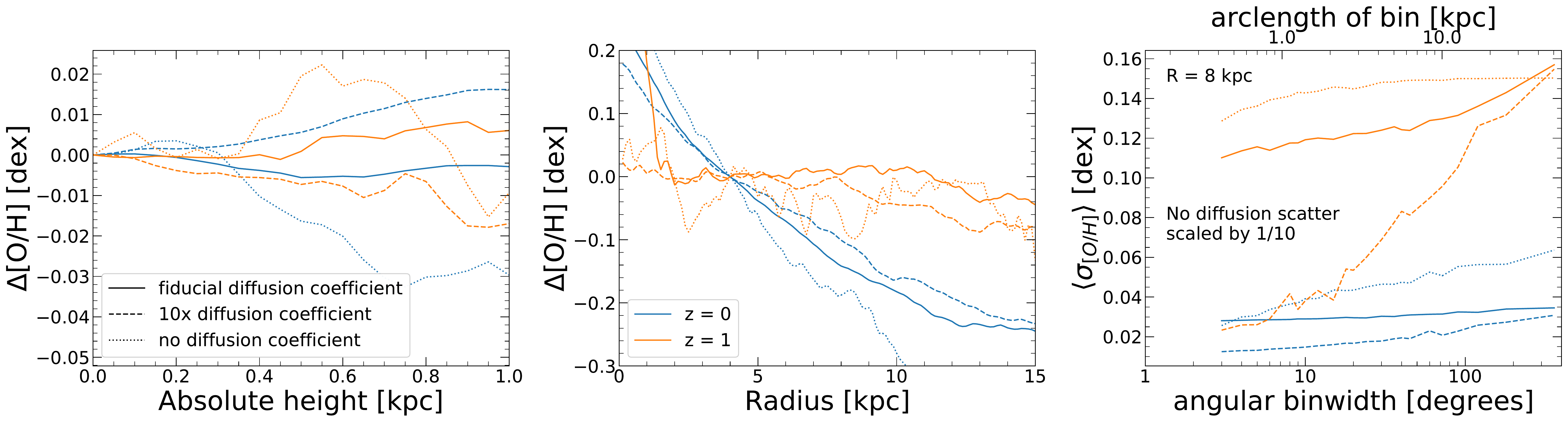}
    \vspace{-5 mm}
    \caption{
    Vertical (left), radial (middle), and azimuthal variations (right) in \OH{} between our fiducial simulation of m12i, a version with no subgrid metal diffusion, and a re-simulation increasing the diffusion coefficient by $10$ times. The vertical profiles show no clear systematic variations at a level important for our analysis. We normalize the radial profiles to $4 \kpc$ (the approximate edge of the bulge) for clarity in comparison. The radial gradients (measured from $4 - 12 \kpc$ for $z = 0$, $2 - 8 \kpc$ for $z = 1$) vary by no more than $\approx 0.005 \dpk$ between our fiducial simulation and the simulation with $10$ times higher metal diffusion, while the simulation with no metal diffusion has a $\approx 0.13 \dpk$ steepe gradient at $z = 0$. In the right panel, we scaled down the azimuthal scatter in the simulation with no metal diffusion by a factor of $10$, for comparison. Thus, neglecting metal diffusion/mixing leads to $10 \times$ higher azimuthal scatter, and moreover, this scatter does not depend much on azimuthal scale. The enhanced metal diffusion re-simulation shows smaller azimuthal scatter at small azimuthal scales, given the enhanced mixing rate on these small scales.
    However, that simulation shows similar scatter at large azimuthal scales, indicating that disk-wide azimuthal scatter is not sensitive to the detailed choice of diffusion coefficient.
    }
    \label{fig:md10_vs_fiducial}
\end{figure*}

Our FIRE-2 simulations model the sub-grid diffusion/mixing of metals in gas via unresolved turbulent eddies \citep{Su17, Escala18, Hopkins18}:
\begin{equation}
    \frac{\partial Z_{i}}{\partial t} + \nabla \cdot (D\nabla Z_{i}) = 0
\end{equation}
where $Z_{i}$ is the mass fraction of a metal in gas element $i$, and $D$ is the diffusion coefficient.
While there is some uncertainty in the exact value to choose for this coefficient, our fiducial value is physically motivated based on tests of the metal diffusion implementation in FIRE-2 on idealized, converged turbulent box simulations by \citet{Colbrook17} and other more extensive studies by \citet{Rennehan19}.
Here, we compare our key results using our fiducial diffusion coefficient $D$ in m12i against a re-simulations of m12i with all identical physics/parameters, except one has a diffusion coefficient that is $10$ times higher (that is, faster mixing) and the other includes no subgrid mixing.

Fig.~\ref{fig:md10_vs_fiducial} compares the vertical, radial, and azimuthal variations for m12i.
The left panel shows the vertical gradient in gas \OH{}, similar to Fig.~\ref{fig:vertical_evolution}.
At $z = 0$, we find no differences within $200 \pc$ and at most $\sim 0.015 \dex$ difference at $1 \kpc$.
The differences are stronger at $z = 1$ for  $10 \times$ higher diffusion and stronger at $z = 0$ for the simulation with no diffusion, though again, not at a significant level to change our interpretations, especially within $200 \pc$.

Fig.~\ref{fig:md10_vs_fiducial} (center) shows the radial profile in gas \OH{}, normalized to the abundance at $R = 4 \kpc$ (given a strong upturn at smaller $R$). The radial gradients, measured over our fiducial radial ranges,
vary by $\lesssim 0.005 \dpk$ between the $10 \times$ diffusion simulation and the fiducial simulation. The gradients vary by less than $0.014 \dex$ between the fiducial simulation and the one with no metal diffusion at $z = 0$. The simulation with no subgrid diffusion has a steeper gradient at $z = 0$, potentially because the metals are less efficient at spreading from a given radius, in the absence of subgrid diffusion, once the disk has become rotationally dominated and radial turbulence is no longer efficient at moving the gas particles.
We thus conclude that the radial gradients are reasonably robust to choices of the strength of the diffusion coefficient, however, in the unphysical case of no subgrid diffusion, the gradient can be (unphysically) steeper.

Fig.~\ref{fig:md10_vs_fiducial} (right) compares the azimuthal variations versus angular bin width, at $R = 8 \kpc$. The simulation with no subgrid diffusion has $10 \times$ higher azimuthal scatter, so we scale down its values in Fig.~\ref{fig:md10_vs_fiducial} by $10 \times$ for visual comparison.  Using no subgrid diffusion leads to scatter that is largely independent of azimuthal scale at high $z$, but that increases with azimuthal scale at low $z$. Without subgrid diffusion, a small number of gas elements can absorb most of the metals, while a significant number of (neighboring) elements can remain nearly un-enriched. This is a patently unphysical scenario, and it yields azimuthal scatter that disagrees with observations by an order of magnitude.
At both redshifts, using a higher diffusion coefficient leads to smaller azimuthal variations at small scales, because diffusion smooths variations between nearby gas elements on scales approaching the resolution \citep{Escala18}.
However, the azimuthal variations are nearly unchanged on large azimuthal scales.
Therefore, our results on small azimuthal scales are likely sensitive to the exact choice of diffusion coefficient, but the large-scale azimuthal variations are robust.  
An important caveat to this comparison is that it is only one simulated galaxy, and individual simulations with the same initial conditions and physics can show non-trivial stochastic variations from random number generators, floating-point roundoff, and chaotic behavior \citep[e.g.][]{Keller19}.
Indeed, we find minor fluctuations between these simulations, for example, in the exact timing of mergers, which can affect all panels in Fig.~\ref{fig:md10_vs_fiducial}.
We consider it likely that the differences in azimuthal variations on small scales are robust, but any other differences in Fig.~\ref{fig:md10_vs_fiducial} are potentially stochastic.


\bsp	
\label{lastpage}
\end{document}